\documentclass[letterpaper]{article} 
\PassOptionsToPackage{table}{xcolor}
\usepackage{aaai2027}
\usepackage[hyphens]{url}
\usepackage{graphicx}
\usepackage{natbib}
\usepackage{caption}
\usepackage{latexsym}
\usepackage{microtype}
\usepackage{enumitem}
\usepackage{pifont}
\usepackage{booktabs}
\usepackage{makecell}
\usepackage{multirow}
\usepackage{tabularx}
\usepackage{array}
\usepackage{adjustbox}
\usepackage{amsmath,amssymb}
\usepackage[most]{tcolorbox}
\usepackage{listings}
\tcbuselibrary{skins,breakable,listings}

\definecolor{PromptGreen}{HTML}{82C95C}
\definecolor{PromptPurple}{HTML}{7A74C9}
\definecolor{PromptBlue}{HTML}{5B8FD9}
\definecolor{PromptOrange}{HTML}{E6A14A}
\definecolor{PromptRed}{HTML}{D96B6B}

\newtcblisting{promptblock}[3][]{%
  enhanced,
  breakable,
  listing only,
  colback=white,
  colframe=#3,
  colbacktitle=#3,
  coltitle=white,
  fonttitle=\bfseries\Large,
  title={#2},
  boxrule=4pt,
  titlerule=0pt,
  arc=0pt,
  outer arc=0pt,
  left=4mm,
  right=4mm,
  top=3mm,
  bottom=3mm,
  before skip=10pt,
  after skip=12pt,
  drop shadow southeast,
  listing options={
    basicstyle=\ttfamily\small,
    columns=fullflexible,
    keepspaces=true,
    breaklines=true,
    breakatwhitespace=false,
    showstringspaces=false,
    tabsize=2
  },
  #1
}

\definecolor{TabHead}{HTML}{D7E3DA}
\definecolor{TabStripe}{HTML}{F2F2F2}
\definecolor{OursStripe}{HTML}{ECECEC}
\definecolor{DropTeal}{HTML}{00A6B2}
\definecolor{GainRed}{HTML}{F05A3D}
\definecolor{MissGray}{HTML}{777777}

\newcommand{\NA}{--}

\newcommand{\modelname}[1]{\texttt{#1}}

\newcommand{\pctup}[1]{#1\%\,\(\uparrow\)}
\newcommand{\pctdown}[1]{#1\%\,\(\downarrow\)}

\newcommand{\tdrop}[1]{{\color{DropTeal}\scriptsize$\downarrow$#1}}

\newcommand{\runner}[1]{\underline{#1}}
\newcommand{\gain}[1]{\hspace{1pt}{\color{GainRed}\scriptsize$\uparrow$#1}}
\DeclareRobustCommand{\helena}{\mbox{\textbf{\texttt{HELENA}}}}

\newcolumntype{Y}{>{\raggedright\arraybackslash}X}

\newcounter{helenaalg}
\newcounter{algline}

\title{HELENA: Hierarchical Sparse Coordination over a Union of Complementary Topologies for MAS}
\author{
Zhifang Mao\textsuperscript{\rm 1, 3},
Linyao Zheng\textsuperscript{\rm 1,2, *},
Xuhang Shi\textsuperscript{\rm 1, 3},
Xiuquan Hou\textsuperscript{\rm 2}
}

\affiliations{
\textsuperscript{\rm 1}XiaoLab\\
\textsuperscript{\rm 2}Xi'an Jiaotong University\\
\textsuperscript{\rm 3}Beijing University of Posts and Telecommunications
\textsuperscript{\rm *}Corresponding author
}

\begin{document}
\maketitle

\begin{abstract}
LLM-based multi-agent systems (MAS) typically optimize a single topology,
restricting reasoning to a narrow trajectory and limiting comprehensive
analytical capacity.
Naively merging multiple topologies into a composite graph introduces redundant noise propagation across irrelevant connections, degrading solution quality.
To address this dilemma, we propose
\textbf{Hierarchical Sparse Coordination over a Union of Complementary Topologies for MAS (HELENA)}, a multi-agent framework that balances
diverse reasoning paths with sparse task-dependent execution.
\helena{} constructs a union MAS graph from complementary candidate topologies selected via Monte Carlo Tree Search and Determinantal Point Process, broadening the reasoning trajectory for comprehensive analysis of complex problems. A Hierarchical Sparse Coordination module then activates only a sparse subgraph at each step while agents exchange compressed latent briefs to suppress redundant noise propagation. Finally, a Local Self-Refinement stage identifies decision units with discrepancy evidence and rewrites them only when contrastive evidence simultaneously confirms a reliable solution-side failure and a challenger-side improvement.
Experiments across eight benchmarks show that \helena{} achieves state-of-the-art results on all benchmarks, with an average gain of \pctup{3.47} over the strongest baseline and up to \pctup{10.34} on MMLU-Pro, achieving larger improvements on harder benchmarks at a reasonable additional cost.

\end{abstract}

\section{Introduction}

LLM-based Multi Agents System(MAS), which integrate language generation with decision-making and action-execution, have shown impressive performance across tasks ranging from reasoning and code generation to video gaming \citep{park2026orak} and autonomous driving \citep{liu2025colmdriver}. More importantly, combining multiple agents into a collaborative team consistently outperforms individuals on complex tasks \citep{wan2026dawn, yang2026birouter}. 
MAS can thus exhibit collective intelligence shaped by collaboration topology and information sharing.

\begin{figure}[t]
  \centering
  \includegraphics[width=\linewidth]{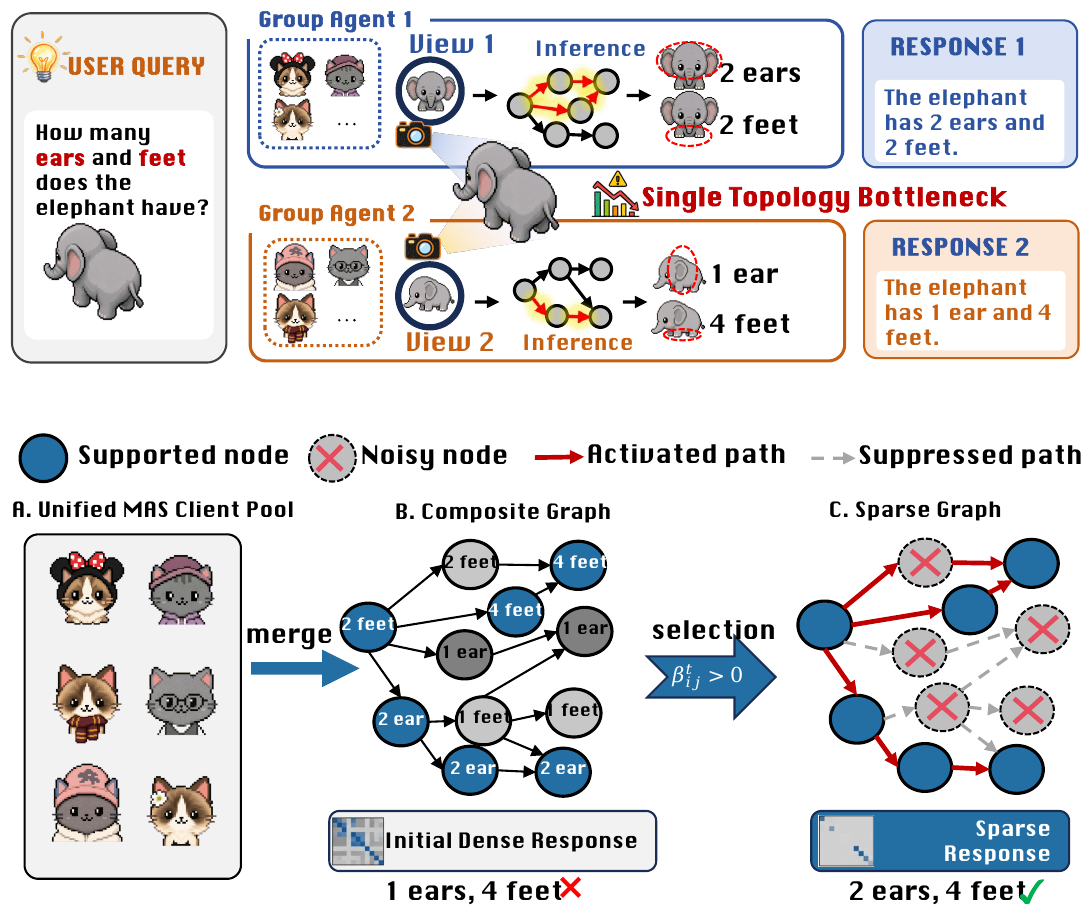}
\caption{
  Problem Illustration: \textbf{I)} A single 
topology restricts reasoning to a narrow trajectory. \textbf{II)} Composite graph propagates 
irrelevant noise across redundant connections.
  }
\label{fig:intro}
\end{figure}

Despite these advances, existing methods for automated multi-agent workflow construction typically optimize a single topology, as in MetaGPT~\citep{hong2024metagpt}, GPTSwarm~\citep{zhuge2024gptswarm}, and G-designer~\citep{zhang2025gdesigner}. This paradigm faces a fundamental bottleneck due to its reliance on a single topology, leading to a narrow reasoning trajectory and preventing a comprehensive analysis of complex problems.
A heuristic idea is to merge multiple topologies into a composite graph.
Rank-and-fuse \citep{jiang2023llmblender}, consensus \citep{chen2024reconcile}, and mixture-style systems confirm that aggregating multiple candidates improves robustness \citep{wang2025mixture}.  
However, a composite topology can propagate irrelevant reasoning paths and feedback signals across redundant connections, which introduces noise into the final solution \citep{huang2025resilience}. As shown in Figure~\ref{fig:intro}, this raises a critical yet under-explored dilemma: 
\textbf{\textit{How to maintain composite reasoning perspectives in MAS while mitigating the propagation of irrelevant noise?}}

A key insight is that fusing complementary topologies captures 
diverse reasoning perspectives, while restricting irrelevant 
node communication prevents noise propagation.
To address this dilemma, we propose \helena{}.
Targeting the first challenge, \ding{172} \textbf{\textit{how to maintain composite reasoning perspectives in MAS}}, \helena{} employs Monte Carlo Tree Search to explore the topology space and selects a complementary subset with a Determinantal Point Process (DPP), merging them into a union graph that structurally preserves diverse reasoning paths.
To further tackle \ding{173} \textbf{\textit{how to mitigate noise propagation induced by composite-graph communication}}, \helena{} addresses this challenge at two levels: by employing a Hierarchical Sparse Coordination module that activates a sparse subgraph at each step while exchanging compressed latent briefs, and by introducing a Local Self-Refinement stage that selects reliable solutions, identifies high-risk units, and revises them based on clear evidence of flaws and alternative improvements.

\noindent\textbf{Our contributions are summarized as follows:}

\begin{itemize}[
    labelindent=-0.35em,  
    labelwidth=0.85em,   
    labelsep=0.35em,     
    leftmargin=!,         
    align=left,
    itemindent=0pt,
    itemsep=0.28em,
    topsep=0.35em,
    parsep=0pt
]
    \item[\ding{182}] \textbf{Dilemma Identification.}
    We identify a topological diversity and noise propagation dilemma in LLM-based MAS. A single topology restricts reasoning coverage, while composite topologies lead to the propagation of irrelevant noise.

    \item[\ding{183}] \textbf{Practical Solution.}
    We propose \helena{}, which builds a union graph from reliable and
    complementary topologies, then controls noise propagation during
    inference through Hierarchical Sparse Coordination and Local
    Self-Refinement.

    \item[\ding{184}] \textbf{Experimental Evaluation.}
Experiments on eight benchmarks show that \helena{} achieves
    state-of-the-art performance, with gains of up to \pctup{10.34}. It also
    maintains favorable cost efficiency across model backbones.
\end{itemize}

\section{Related Work}

\paragraph{Pre-defined MAS Protocols.}
Early LLM-based multi-agent systems rely on pre-defined collaboration or aggregation protocols \citep{wu2023autogen}, including role-based systems that assign predefined responsibilities \citep{li2023camel, chen2024agentverse, hong2024metagpt}, debate-style methods that encourage agents to challenge intermediate reasoning \citep{du2024improving, liang2024encouraging}, and rank-and-fuse approaches such as LLM-Blender \citep{jiang2023llmblender}. While these methods improve over individual LLMs, their manually specified and task-agnostic protocols limit adaptability across domains and resource budgets.

\paragraph{Automated Agentic System.}
To reduce manual engineering, recent work formulates agentic construction as automated optimization. AFlow and A$^2$Flow search over code-represented workflows \citep{zhang2025aflow, zhao2026a2flow}, ADAS and AgentSquare expand the search space to modular designs \citep{hu2025automated, shang2025agentsquare}, and MaAS optimizes a supernet for query-dependent architecture sampling \citep{zhang2025maas}. These methods show that topology can be searched rather than manually specified.

\paragraph{Multi-agent Systems as Graphs.}
GPTSwarm represents agents as optimizable graphs \citep{zhuge2024gptswarm}, G-Designer learns task-aware topologies via graph neural networks \citep{zhang2025gdesigner}, MacNet studies scalable graph-organized collaboration \citep{qian2025scaling}, and G-Memory organizes memory traces with graph structures \citep{zhang2025gmemory}. However, both automated optimization and graph-based methods predominantly instantiate a single topology at a time, leaving open how to preserve complementary topological perspectives without suffering from noise propagation induced by composite-graph communication.



\section{Preliminaries}


We model a multi-agent system as a directed collaboration graph
$G=(V,E)$, where $V=\{v_1,\ldots,v_N\}$ is the set of agent nodes,
$N=|V|$, and $E$ is the set of directed edges. Each node $v_i\in V$
corresponds to an LLM-based agent and is formalized as
\begin{equation}
  v_i = \left(\mathrm{Base}_i,\ \mathrm{Role}_i,\ \mathrm{Mem}_i\right),
  \label{eq:agent}
\end{equation}
where $\mathrm{Base}_i$ denotes the underlying large language model
instance, $\mathrm{Role}_i$ specifies the agent's designated role or
persona, and $\mathrm{Mem}_i$ denotes its memory state.

Given a user query $Q$, the system evolves through $T$ communication
epochs. At epoch $t$, each agent receives the query, its own memory state,
and the information made available by its in-neighbors:
\begin{equation}
\begin{aligned}
  o_i^t &= v_i\left(Q,\mathrm{Mem}_i^t,\mathcal{C}_i^t\right)
\end{aligned}
\label{eq:agent-output}
\end{equation}
Here, $\mathcal{C}_i^t=\{m_{j\to i}^t:v_j\in\mathcal{N}^{-}(v_i)\}$ denotes
the information made available from in-neighbor agents to $v_i$ at epoch $t$, $o_i^t$ denotes the output generated by $v_i$, which may include
reasoning steps, intermediate analyses, or final proposals, and $m_{j\to i}^t$ denotes the message transmitted from agent $v_j$ to agent $v_i$ at epoch $t$.

\section{Method}

\begin{figure*}[t]
  \centering
  \includegraphics[width=\linewidth]{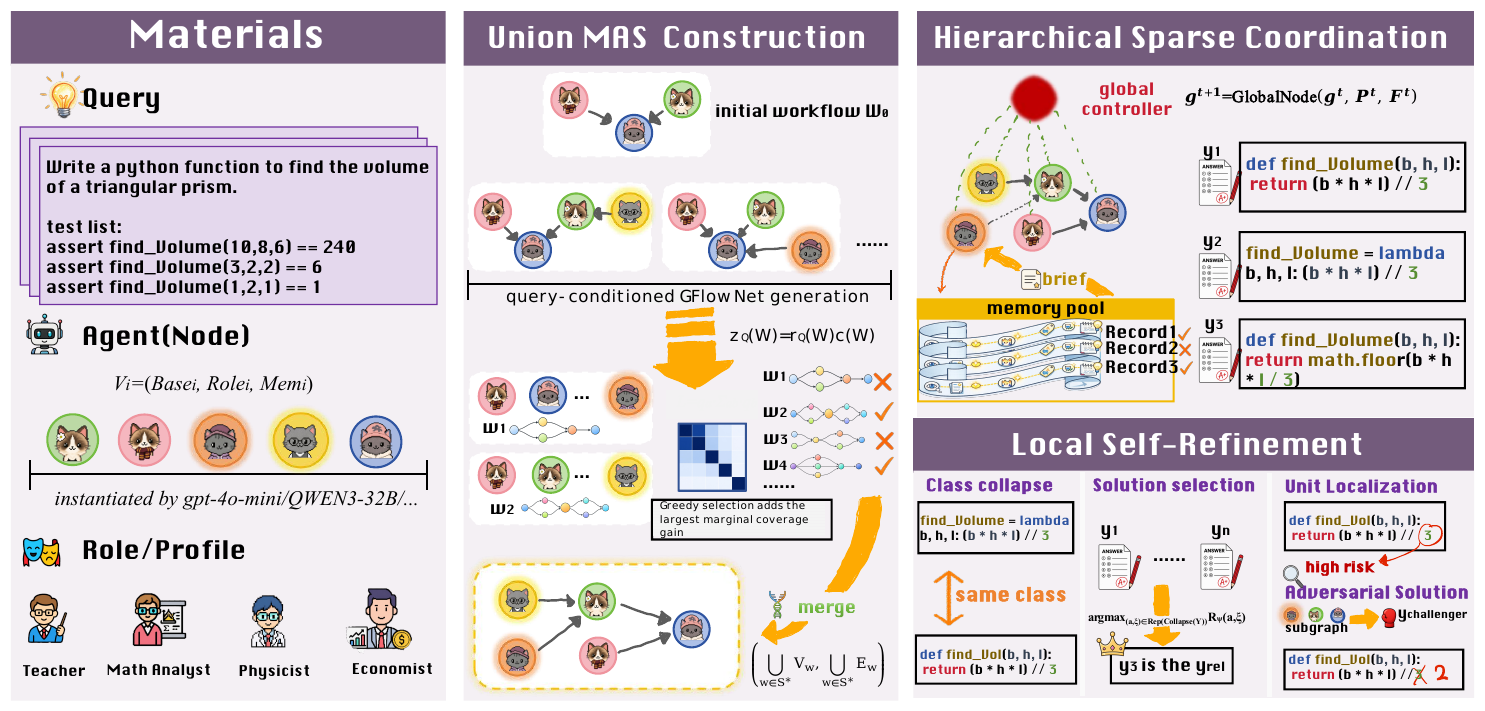}
  \caption{
  The designing workflow of our proposed \helena{}.
  }
\label{fig:method}
\end{figure*}

The overall framework of \helena{} is illustrated in Figure~\ref{fig:method}.

\subsection{Union MAS Graph Construction}
\label{sec:union-graph}

\helena{} generates candidate topologies for the query and then selects a
reliable and complementary subset to construct a query-dependent union graph.

\paragraph{Candidate Topology Generation.}
Given a query $Q$ and an initial workflow $W_0$, candidate topologies are
generated through legal graph edits. At search state $s$, an edit $u$ changes
an agent role, a prompt setting, or a directed edge while preserving workflow
validity. Tree search has recently been used to navigate discrete agent workflow spaces
\citep{zhang2025aflow,li2026agentswift}. In \helena{}, topology search is
instantiated as a query-conditioned variant of policy-value MCTS
    \citep{silver2017mastering,zhao2022efficient}.
The edit prior $P_\phi(u\mid s,Q)$ guides action selection, while the utility
predictor $g_\psi$ evaluates terminal topologies. The valid terminal topologies
found within search budget $B$ constitute the query-specific candidate pool
$\mathcal C(Q)$.

During inference, the search is reinitialized for each query. Both
$P_\phi$ and $g_\psi$ remain frozen, and candidate workflows are not executed.
Agent LLM calls begin only after the selected topologies have been merged into
$G_U$. The offline learning procedure is provided in
Appendix~\ref{app:algorithm-workflow}.

\paragraph{Union Graph Construction.}
Candidate selection is based on query-dependent quality and information-flow
similarity. For each candidate topology $W$, the quality predictor $g_\psi$
takes $Q$ and $W$ as input and returns the expected utility $\mu_{Q,W}$ and
predictive uncertainty $\sigma_{Q,W}$. The lower quantile is denoted by
$\alpha<0.5$, and $\Phi^{-1}$ is the inverse cumulative distribution function
of the standard normal distribution. The parameter $\eta>0$ controls the scale
of the quality weight. The query-dependent quality weight is defined as
\begin{equation}
q_{Q,W}
=
\exp\left(
\frac{
\mu_{Q,W}+\Phi^{-1}(\alpha)\sigma_{Q,W}
}{
2\eta
}
\right).
\label{eq:query-quality}
\end{equation}

The weight increases with predicted utility and decreases with predictive
uncertainty. Its dependence on both $Q$ and $W$ allows the same topology to
receive different weights for different queries.

Quality alone cannot determine whether two candidates provide distinct
coordination patterns. Their complementarity is therefore measured from the
information flows supported within $T$ communication turns. After
role-compatible nodes are mapped to a common order, $A_W$ denotes the
normalized adjacency matrix of topology $W$. Its finite-turn information flow
is
\begin{equation}
\Psi_T(W)
=
\sum_{t=1}^{T}A_W^t.
\label{eq:information-flow}
\end{equation}

The matrix power $A_W^t$ records the directed routes that can transmit
information in $t$ steps. Their finite-turn coverage is represented by
$\Psi_T(W)$.

Matrix vectorization is denoted by $\operatorname{vec}(\cdot)$. For two
candidate topologies $W_a$ and $W_b$, their information-flow similarity is
\begin{equation}
\Sigma_{ab}
=
\frac{
\left\langle
\operatorname{vec}\left(\Psi_T(W_a)\right),
\operatorname{vec}\left(\Psi_T(W_b)\right)
\right\rangle
}{
\left\|
\operatorname{vec}\left(\Psi_T(W_a)\right)
\right\|_2
\left\|
\operatorname{vec}\left(\Psi_T(W_b)\right)
\right\|_2
}.
\label{eq:information-flow-similarity}
\end{equation}

A larger $\Sigma_{ab}$ indicates greater overlap between the directed
information flows of the two topologies. Query-dependent quality and
information-flow similarity define the DPP kernel
\begin{equation}
L_{ab}
=
q_{Q,W_a}\Sigma_{ab}q_{Q,W_b}.
\label{eq:dpp-kernel}
\end{equation}

The selected topology set is defined as
\begin{equation}
S^\star
=
\operatorname*{arg\,max}_{S\subseteq\mathcal C(Q),\,|S|=K}
\log\det(L_S).
\label{eq:dpp-selection}
\end{equation}

Here $K$ is the number of retained topologies, and $L_S$ is the principal
submatrix indexed by $S$. The log-determinant criterion assigns high values to
sets that combine predicted quality with nonredundant information flows.
Greedy MAP inference is used to obtain $S^\star$.

Before merging, nodes with the same role and agent identity are aligned.
Different prompt settings are retained as variants of the aligned node.
The node and edge sets of topology $W$ are denoted by $V_W$ and $E_W$.
The selected topologies form the union graph
\begin{equation}
G_U
=
\left(
\bigcup_{W\in S^\star}V_W,
\bigcup_{W\in S^\star}E_W
\right).
\label{eq:union-graph}
\end{equation}

The union retains the directed edges of the selected topologies but does not
represent the strength of their support. Predictive confidence is therefore
assigned to each union edge $u\rightarrow v$ as
\begin{equation}
p_{uv}
=
\frac{
\sum_{W\in S^\star}
q_{Q,W}\mathbf 1\left[(u,v)\in E_W\right]
}{
\sum_{W\in S^\star}q_{Q,W}
}.
\label{eq:edge-confidence}
\end{equation}

Here $\mathbf 1[(u,v)\in E_W]$ equals one when $W$ contains the edge and zero
otherwise. Support from a topology with a larger quality weight contributes
more to $p_{uv}$. Each edge descriptor is written as
\begin{equation}
e_{u\rightarrow v}
=
\left(
p_{uv},
\mathrm{compat}_{uv},
\mathrm{ctx}_{uv}
\right).
\label{eq:edge-descriptor}
\end{equation}
The remaining components represent endpoint compatibility and source topology
context. Their construction is provided in
Appendix~\ref{app:algorithm-workflow}.

\subsection{Hierarchical Sparse Coordination}

Hierarchical Sparse Coordination realizes sparse collaboration at two levels.
Node-Level Memory Composer removes irrelevant private records before they enter
coordination. Edge-Level Sparse Activation restricts communication and
execution to selected paths at each turn.

\paragraph{Node-Level Memory Composer.}
Each node $v_i$ maintains a private memory store
$\mathcal M_i^t=\{m_{ik}^t\}$. Let $r_i$ denote its role instruction. The query
builder $f_q$ forms the retrieval feature $q_i^t=f_q(Q,r_i)$. The relevance
scorer $f_s$ assigns each memory record the score
$z_{ik}^t=f_s(q_i^t,m_{ik}^t)$. Both modules are trained from retrospective
execution feedback as described in Appendix~\ref{app:algorithm-workflow}.

Sparsemax \citep{martins2016sparsemax}  converts these scores into the retrieval weight vector
\begin{equation}
\boldsymbol\alpha_i^t
=
\operatorname{sparsemax}
\left(
\{z_{ik}^t\mid m_{ik}^t\in\mathcal M_i^t\}
\right).
\label{eq:memory-sparsemax}
\end{equation}

Its component $\alpha_{ik}^t$ is the weight of record $m_{ik}^t$. Records with
positive weight form the selected set $\mathcal R_i^t$.

The composer returns the private context $\ell_i^t$. The selected records are
also represented by a latent brief $b_i^t$ and a verbalized brief $B_i^t$.
\begin{equation}
(\ell_i^t,b_i^t,B_i^t)
=
\operatorname{Compose}
\left(
\{(\alpha_{ik}^t,m_{ik}^t)\mid m_{ik}^t\in\mathcal R_i^t\}
\right).
\label{eq:memory-compose}
\end{equation}

This node-level sparsity removes irrelevant private records before
coordination. The selected memory remains private in $\ell_i^t$, while $b_i^t$
supports graph control and $B_i^t$ is transmitted across active edges.

\paragraph{Edge-Level Sparse Activation.}
Let $g^t$ denote the global state at turn $t$. Let $\Theta_e$ denote the
parameters of the edge scorer. For each edge $(v_j,v_i)\in E_U$, the scorer
combines $g^t$ with an edge representation formed from the endpoint briefs and
the descriptor defined in Section~\ref{sec:union-graph}.
\begin{equation}
\phi_{ji}^t
=
\Psi_{\Theta_e}^{\mathrm{edge}}
\left(
b_j^t,b_i^t,g^t,e_{v_j\rightarrow v_i}
\right).
\label{eq:edge-score}
\end{equation}

For each target node $v_i$, sparsemax produces the incoming edge weight vector
\begin{equation}
\boldsymbol\beta_i^t
=
\operatorname{sparsemax}
\left(
\{\phi_{ji}^t\mid(v_j,v_i)\in E_U\}
\right).
\label{eq:edge-sparsemax}
\end{equation}

Its component $\beta_{ji}^t$ is the weight assigned to edge $(v_j,v_i)$. The
active edge set is
\begin{equation}
E^t
=
\{(v_j,v_i)\in E_U\mid\beta_{ji}^t>0\}.
\label{eq:active-edges}
\end{equation}

The endpoints of $E^t$ form $V^t$, and $G^t=(V^t,E^t)$ is executed at turn
$t$. This edge-level sparsity restricts communication to $E^t$ and agent execution
to $V^t$, so coordination proceeds only along the selected paths.

Let $D_i^t$ denote the predecessor context formed from the verbalized briefs
received through active incoming edges. Let $P_i^t$ denote the role-conditioned
prompt of $v_i$. Each active Agent produces
\begin{equation}
o_i^t
=
v_i
\left(
P_i^t,Q,\ell_i^t,D_i^t,g^t
\right).
\label{eq:generation}
\end{equation}

The active briefs and outputs are pooled into $p^t$. Feedback available during
inference forms $\mathcal F^t$ and does not contain benchmark labels. Let
$\Theta_g$ denote the parameters of the global updater. The next state is
\begin{equation}
g^{t+1}
=
\operatorname{GlobalNode}_{\Theta_g}
\left(
g^t,p^t,\mathcal F^t
\right).
\label{eq:global-update}
\end{equation}

The updated state conditions edge selection at the next turn, allowing the
active graph to change with the current execution state. The edge scorer and global updater are trained offline and remain frozen during
evaluation. When $o_i^t$ is a solution proposal, it is added to
$Y=\{(a_k,\xi_k)\}_{k=1}^{K_y}$, where $K_y=|Y|$. Each $\xi_k$ links its answer
to the active graph trace and the evidence supporting that answer. The active
graphs are recorded as
$G_{\mathrm{act}}^{1:T}=(G^1,\ldots,G^T)$.

\subsection{Local Self-Refinement}

Residual errors can persist after Hierarchical Sparse Coordination. Local
Self-Refinement addresses them through Decision Unit Localization and
Adversarial Solution Validation.

\paragraph{Decision Unit Localization.}
Hierarchical Sparse Coordination produces the solution set
$Y=\{(a_k,\xi_k)\}_{k=1}^{K_y}$. Each candidate answer $a_k$ is paired with an
evidence record $\xi_k$ that links it to the activated graph trace and its
supporting evidence. Each record $\xi_k$ links an answer to the activated
graph trace and the evidence that supports it. Following semantic answer
equivalence \citep{kuhn2023semantic}, the candidates are collapsed into answer
classes. The evidence-aware reliability estimator $R_\psi$ scores the candidates
within each class. The selected answer and its evidence record are
\begin{equation}
(y_{\mathrm{rel}},\xi_{\mathrm{rel}})
=
\operatorname*{arg\,max}_{(a,\xi)\in
\operatorname{Rep}(\operatorname{Collapse}(Y))}
R_\psi(a,\xi).
\label{eq-reliable-selection}
\end{equation}
Here, $\operatorname{Collapse}(Y)$ groups semantically equivalent answers and
$\operatorname{Rep}$ retains the highest-scoring member of each class. The
selected answer serves as the default solution and remains unchanged unless
local validation supports a replacement.

The selected solution is decomposed into locally verifiable decision units. For
each unit $d$, the discrepancy encoder $\mathrm{Disc}_\theta$ maps the evidence
linked to $d$ into the representation $\epsilon_d$. This representation
indicates whether the unit has consistent support. The risk scorer
$\mathrm{risk}_\theta$ maps $\epsilon_d$ to a scalar value. Let $\tau$ denote
the risk threshold. The units selected for further validation are
\begin{equation}
\mathcal U
=
\left\{
d\in\operatorname{UnitParse}(y_{\mathrm{rel}},\xi_{\mathrm{rel}})
\mid
\mathrm{risk}_\theta(d,\epsilon_d)>\tau
\right\}.
\label{eq-risky-loci}
\end{equation}
Units outside $\mathcal U$ are kept unchanged. The task-specific units and the
local refinement operators are defined in Appendix~A.2. This stage uses only
evidence available during inference and does not access benchmark-only
supervision.

\paragraph{Adversarial Solution Validation.}
For each unit $d\in\mathcal U$, a local correction subgraph is extracted from
$G_U$ by following the active evidence paths stored in $\xi_{\mathrm{rel}}$.
The subgraph retains the nodes that contributed to or checked the disputed unit.
A challenger $y_d^{\mathrm{ch}}$ with evidence record $\xi_d^{\mathrm{ch}}$ is
retrieved from $Y$ when another supported candidate disagrees at $d$. Otherwise,
it is generated by executing the correction subgraph with a targeted repair
prompt. The term adversarial means that the challenger contests one local
decision.

The unit-conditioned score $R_\psi^{\mathrm{loc}}$ applies the same reliability
estimator to evidence linked to $d$. The local gain $\Delta_d$ is the score of
the challenger minus the score of the selected solution. Let $\delta$ denote
the required improvement margin. The certificate model compares the proposed
replacement with the current unit using their linked evidence. Its output is
\begin{equation}
c_d
=
\mathrm{Cert}_\chi
\left(
y_d^{\mathrm{ch}},
\xi_d^{\mathrm{ch}},
y_{\mathrm{rel}},
\xi_{\mathrm{rel}},
d
\right).
\label{eq-certificate-output}
\end{equation}
It equals one when the evidence supports the replacement while preserving the
remaining solution. Let $\mathbb I$ denote the indicator function. The
acceptance rule is
\begin{equation}
\mathrm{Accept}(y_d^{\mathrm{ch}},d)
=
\mathbb I[\Delta_d>\delta\land c_d=1].
\label{eq-certificate-gating}
\end{equation}
When a challenger is accepted, only unit $d$ is replaced and the evidence record
is updated. If no challenger satisfies the acceptance rule, the selected
solution is preserved.

\section{Experiments}

\subsection{Experiment Setup}

\paragraph{Datasets.}
We evaluate \helena{} on eight widely used benchmarks covering three representative domains.
For general knowledge and professional reasoning, we use \textbf{MMLU}~\citep{hendrycks2021mmlu} and \textbf{MMLU-Pro}~\citep{wang2024mmlupro}.
For mathematical reasoning, we use \textbf{GSM8K}~\citep{cobbe2021gsm8k}, \textbf{MATH}~\citep{hendrycks2021math}, \textbf{MATH-Lv5} ~\citep{hendrycks2021math}, and \textbf{SVAMP}~\citep{patel2021svamp}.
For code generation, we use \textbf{HumanEval}~\citep{chen2021codex} and \textbf{MBPP}~\citep{austin2021program}.

\paragraph{Baselines.}


We compare \helena{} with representative single-agent, protocol-based,
automated agentic, and graph-structured MAS baselines listed in
Table~\ref{tab:main-results}. All results are independently reproduced under
the same evaluation protocol using official implementations when available.
Implementation details and budget controls are provided in Appendix~A.1.

\paragraph{Implementation Details.}
All agents use \modelname{gpt-4o-mini}; topology and memory embeddings use
\modelname{Qwen3-Embedding-8B} with $D=512$. Local Qwen models run on eight
NVIDIA H100 GPUs. We use $K=3$ fused topologies and average three independent
runs. Code tasks are evaluated by \textit{pass@1}; all other tasks use
exact-match accuracy. Additional settings are provided in the supplementary material.

\paragraph{Evaluation Protocol.}
Observed topology utilities are computed only from training examples.
Validation labels are used only for model selection and calibration after each
system output is fixed. During test-time MCTS and topology selection, the
system receives the query and frozen model predictions only; gold answers,
answer indices, exact-match feedback, and official held-out tests are
inaccessible. Prompt-visible examples are treated as part of $Q$ and are distinct
from hidden evaluator tests. The benchmark scorer is invoked only after the
final output has been fixed.
\subsection{Main Results}

\begin{table*}[t]
\centering
\caption{
Performance comparison under the unified reproduction protocol. All baselines are independently reproduced by us using the same \modelname{gpt-4o-mini} backbone, benchmark instances, decoding settings, and evaluation scripts. The best result is shown in bold and the runner-up is underlined. Red arrows denote the absolute improvement over the strongest reproduced baseline.
}
\label{tab:main-results}

\begingroup
\scriptsize
\setlength{\tabcolsep}{4.0pt}
\renewcommand{\arraystretch}{1.08}
\arrayrulecolor{black}

\begin{adjustbox}{max width=\textwidth}
\begin{tabular}{l!{\color{black}\vrule width .45pt}cc!{\color{black}\vrule width .45pt}cccc!{\color{black}\vrule width .45pt}cc}
\specialrule{1.15pt}{0pt}{0pt}

\rowcolor{TabHead}
\textbf{Method}
& \multicolumn{2}{c!{\color{black}\vrule width .45pt}}{\textbf{Code generation}}
& \multicolumn{4}{c!{\color{black}\vrule width .45pt}}{\textbf{Mathematical reasoning}}
& \multicolumn{2}{c}{\textbf{Knowledge / professional}} \\

\rowcolor{TabHead}
& \textbf{MBPP}
& \textbf{HumanEval}
& \textbf{GSM8K}
& \textbf{MATH}
& \textbf{SVAMP}
& \textbf{MATH-Lv5}
& \textbf{MMLU}
& \textbf{MMLU-Pro} \\

\specialrule{0.75pt}{0pt}{0pt}

\multicolumn{9}{l}{\textbf{\textit{Single-Agent}}} \\[-1pt]

Self-Refine~\citep{madaan2023selfrefine}
& 69.80
& 87.80
& 89.60
& 46.10
& 88.73
& 32.85
& 75.44
& 57.97 \\

\specialrule{0.35pt}{0pt}{0pt}

\multicolumn{9}{l}{\textbf{\textit{Pre-defined MAS Protocols}}} \\[-1pt]

\rowcolor{TabStripe}
LLM-Debate~\citep{du2024improving}
& 70.29
& 88.68
& 89.47
& 52.96
& 91.76
& 38.55
& 81.04
& 63.28 \\

AgentVerse~\citep{chen2024agentverse}
& 74.28
& 89.29
& 89.91
& 50.85
& 89.64
& 36.85
& 78.36
& 60.83 \\

\rowcolor{TabStripe}
LLM-Blender~\citep{jiang2023llmblender}
& 77.05
& 88.80
& 88.35
& 50.34
& 89.52
& 36.16
& 81.22
& 63.47 \\

DyLAN~\citep{liu2023dylan}
& 77.30
& 90.42
& 89.98
& 51.12
& 88.48
& 37.38
& 79.96
& 62.14 \\

\specialrule{0.35pt}{0pt}{0pt}

\multicolumn{9}{l}{\textbf{\textit{Automated Agentic Systems}}} \\[-1pt]

\rowcolor{TabStripe}
H-Swarms~\citep{feng2025heterogeneous}
& 81.65
& 89.86
& 95.00
& 65.89
& 92.74
& 50.27
& 83.68
& \runner{69.00} \\

AFlow~\citep{zhang2025aflow}
& 82.20
& 90.06
& 92.30
& 73.35
& 91.73
& 58.82
& 83.10
& 64.35 \\

\rowcolor{TabStripe}
RouterDC~\citep{chen2024routerdc}
& 75.20
& 87.75
& 93.68
& 73.46
& 91.86
& 58.93
& 82.01
& 63.27 \\

A$^2$Flow~\citep{zhao2026a2flow}
& 85.00
& 92.40
& 93.80
& 58.50
& 92.15
& 44.76
& 83.29
& 63.42 \\

\rowcolor{TabStripe}
DAAO~\citep{su2025difficulty}
& \runner{86.95}
& \runner{94.65}
& 94.40
& 55.37
& 92.64
& 42.57
& 84.90
& 65.28 \\

MasRouter~\citep{yue2025masrouter}
& 84.00
& 90.62
& \runner{95.45}
& \runner{75.42}
& 92.95
& \runner{61.36}
& 84.25
& 64.85 \\

\rowcolor{TabStripe}
MaAS~\citep{zhang2025maas}
& 82.17
& 92.85
& 92.30
& 74.45
& 91.55
& 60.23
& 83.01
& 63.68 \\

BiRouter~\citep{yang2026birouter}
& 84.82
& 91.46
& 94.09
& 74.92
& \runner{93.20}
& 60.87
& 86.80
& 66.53 \\

\specialrule{0.35pt}{0pt}{0pt}

\multicolumn{9}{l}{\textbf{\textit{Multi-agent Systems as Graphs}}} \\[-1pt]

\rowcolor{TabStripe}
GPTSwarm~\citep{zhuge2024gptswarm}
& 75.40
& 86.28
& 94.66
& 68.85
& 92.18
& 54.64
& 82.80
& 64.19 \\

G-Designer~\citep{zhang2025gdesigner}
& 79.24
& 87.50
& 93.97
& 70.46
& 90.29
& 55.39
& \runner{87.20}
& 66.94 \\

\rowcolor{TabStripe}
DAWN~\citep{wan2026dawn}
& 86.70
& 94.44
& 93.12
& 73.80
& 92.10
& 59.21
& 80.00
& 62.15 \\

\specialrule{0.75pt}{0pt}{0pt}

\rowcolor{OursStripe}
\textbf{\texttt{\helena{} (Ours)}}
& \textbf{88.73}\gain{1.78}
& \textbf{95.20}\gain{0.55}
& \textbf{97.39}\gain{1.94}
& \textbf{78.29}\gain{2.87}
& \textbf{95.48}\gain{2.28}
& \textbf{68.21}\gain{6.85}
& \textbf{88.37}\gain{1.17}
& \textbf{79.34}\gain{10.34} \\

\specialrule{1.15pt}{0pt}{0pt}
\end{tabular}
\end{adjustbox}
\endgroup

\vspace{0.6mm}
\begin{minipage}{0.985\textwidth}
\footnotesize
\emph{Notes.}
All scores are percentages.
MATH-Lv5 denotes the Level-5 subset of MATH, and MMLU-Pro denotes the harder professional-knowledge variant of MMLU.

\end{minipage}

\end{table*}

\textbf{Obs.\ding{182} State-of-the-Art Performance with Difficulty-Scaled Gains.}
As shown in Table~\ref{tab:main-results}, \helena{} achieves the best results on all evaluated benchmarks, with an average gain of \pctup{3.47} over the strongest baseline. The performance margin consistently widens with task difficulty: \helena{} outperforms H-Swarms by \pctup{10.34} points on MMLU-Pro and exceeds MasRouter by \pctup{6.85} points on MATH-Lv5. These results demonstrate that topology-diverse coordination with noise-controlled execution yields substantial benefits on complex, multi-step reasoning tasks.


\subsection{Cost Analysis}

\begingroup
\setcounter{dbltopnumber}{3}
\renewcommand{\dbltopfraction}{0.95}
\renewcommand{\textfraction}{0.04}
\setcounter{bottomnumber}{1}
\renewcommand{\bottomfraction}{0.72}
\setlength{\dblfloatsep}{5pt plus 1pt minus 1pt}
\setlength{\dbltextfloatsep}{7pt plus 1pt minus 2pt}
\setlength{\floatsep}{5pt plus 1pt minus 1pt}
\setlength{\textfloatsep}{7pt plus 1pt minus 2pt}

\begin{figure*}[t]
  \centering
  \begin{minipage}[t]{0.49\textwidth}
    \centering
    \includegraphics[width=\linewidth]{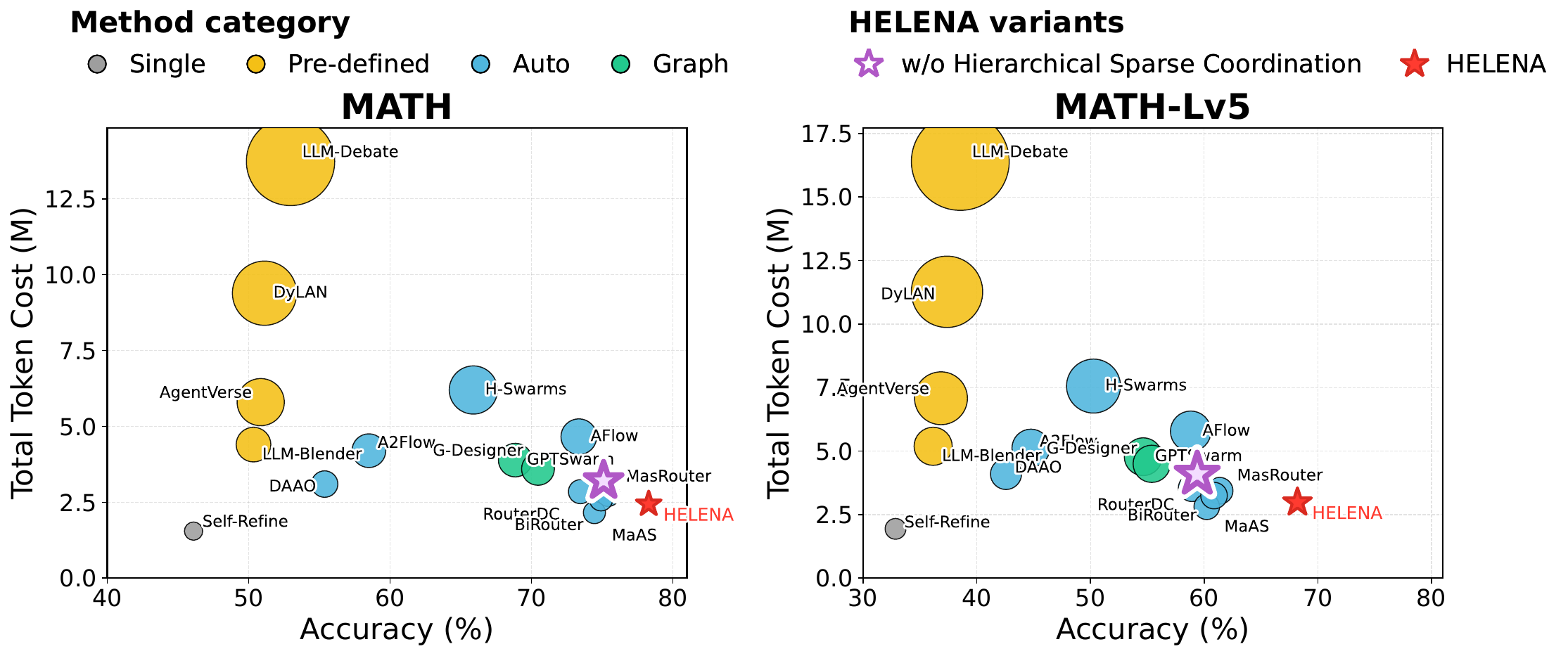}
    \centerline{\small (a) MATH and MATH-Lv5}
  \end{minipage}
  \hfill
  \begin{minipage}[t]{0.49\textwidth}
    \centering
    \includegraphics[width=\linewidth]{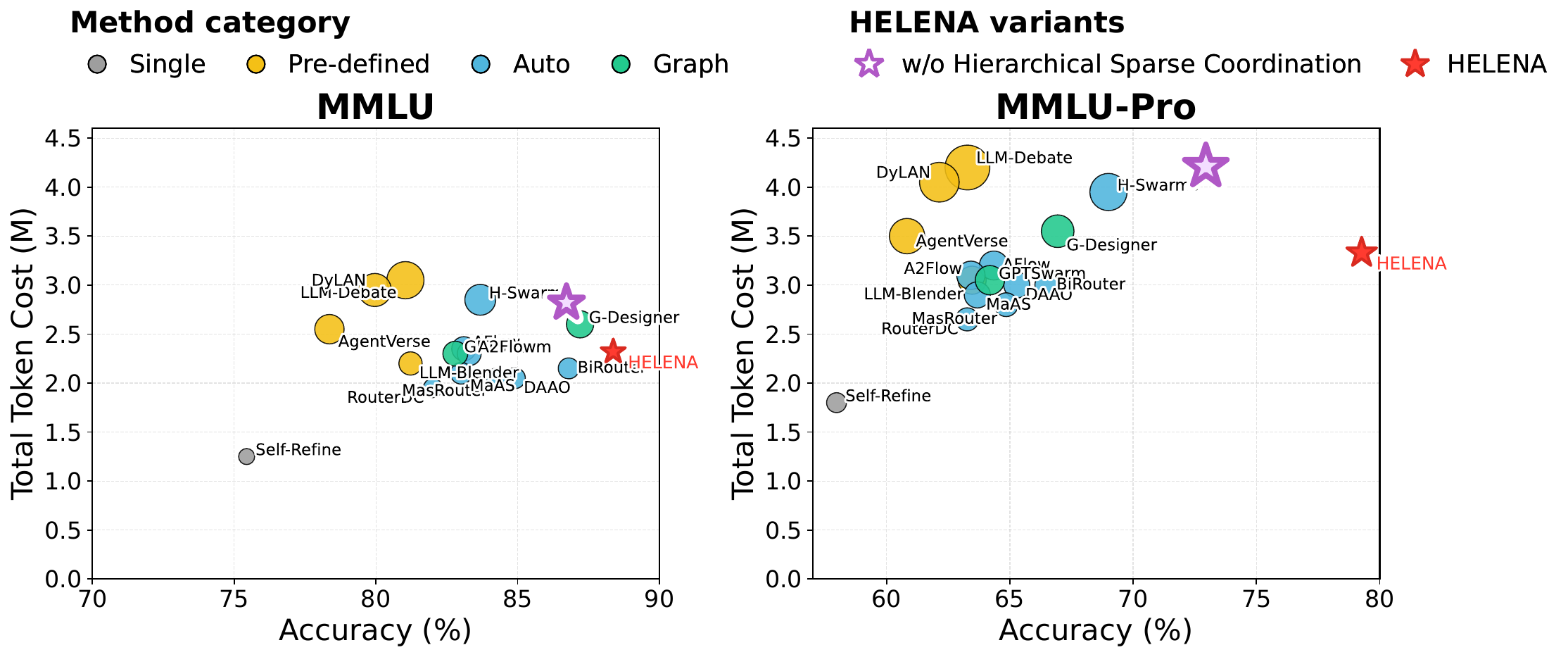}
    \centerline{\small (b) MMLU and MMLU-Pro}
  \end{minipage}

  \caption{
  Cost--accuracy comparison and difficulty-dependent scaling of \helena{}.
  Left: MATH and MATH-Lv5. Right: MMLU and MMLU-Pro.
  }
  \label{fig:cost}
\end{figure*}

\textbf{Obs.\ding{183} Difficulty-Adaptive Token Efficiency.}
Figure~\ref{fig:cost} shows that \helena{} achieves a favorable 
cost--accuracy trade-off across benchmarks of varying difficulty. 
On standard datasets, \helena{} remains in the low-cost region 
(2.44M tokens on MATH, 2.32M tokens on MMLU) while achieving 
the best accuracy. As reasoning complexity increases, \helena{} 
adaptively scales computation: relative cost increases to 
1.25$\times$ on MATH-Lv5 and 1.40$\times$ on MMLU-Pro.

Concurrently, Figure~\ref{fig:cost} reveals that this efficiency primarily 
stems from the Node-Level Memory Composer and Edge-Level Sparse Activation. The memory 
composer selects and compresses task-relevant records into compact 
latent and verbalized briefs, while the controller activates only 
a sparse subgraph over the union graph, preventing full-graph 
communication and redundant context propagation.

\subsection{Ablation Study and Sensitivity Analysis}

We conduct ablations on four key components of \helena{}: local refinement, union graph construction, latent memory, and the global controller.
Table~\ref{tab:ablation} reports the results, where each variant disables one component while keeping the remaining system unchanged.


\begin{table*}[t]
\centering
\caption{Ablation study.}
\label{tab:ablation}
\begingroup
\scriptsize
\setlength{\tabcolsep}{3.8pt}
\renewcommand{\arraystretch}{1.14}
\arrayrulecolor{black}

\begin{adjustbox}{max width=\textwidth}
\begin{tabular}{
l
!{\color{black}\vrule width .45pt}cc
!{\color{black}\vrule width .45pt}cccc
!{\color{black}\vrule width .45pt}cc
!{\color{black}\vrule width .45pt}cc
}
\specialrule{1.10pt}{0pt}{0pt}

\rowcolor{TabHead}
\textbf{Variant}
& \multicolumn{2}{c!{\color{black}\vrule width .45pt}}{\textbf{Code generation}}
& \multicolumn{4}{c!{\color{black}\vrule width .45pt}}{\textbf{Mathematical reasoning}}
& \multicolumn{2}{c!{\color{black}\vrule width .45pt}}{\textbf{Knowledge / professional}}
& \multicolumn{2}{c}{\textbf{Summary}} \\

\rowcolor{TabHead}
& \textbf{MBPP}
& \textbf{HumanEval}
& \textbf{GSM8K}
& \textbf{MATH}
& \textbf{SVAMP}
& \textbf{MATH-Lv5}
& \textbf{MMLU}
& \textbf{MMLU-Pro}
& \textbf{Avg.}
& \(\boldsymbol{\Delta}\)\textbf{Avg.} \\

\specialrule{0.75pt}{0pt}{0pt}

\rowcolor{OursStripe}
\textbf{\helena{} (Ours)}
& \textbf{88.73}
& \textbf{95.20}
& \textbf{97.39}
& \textbf{78.29}
& \textbf{95.48}
& \textbf{68.21}
& \textbf{88.37}
& \textbf{79.34}
& \textbf{86.37}
& \NA \\

\specialrule{0.35pt}{1.5pt}{1.5pt}

\rowcolor{TabStripe}
w/o union graph
& 84.74
& 87.36
& 96.87
& 76.89
& 93.18
& 66.07
& 86.70
& 76.12
& 83.49
& \tdrop{2.88} \\

w/o memory composer
& 84.59
& 88.09
& 96.58
& 76.15
& 94.20
& 63.49
& 87.24
& 75.42
& 83.22
& \tdrop{3.15} \\

\rowcolor{TabStripe}
w/o sparse activation
& 86.15
& 90.77
& 97.04
& 77.24
& 94.83
& 64.11
& 87.85
& 76.80
& 84.35
& \tdrop{2.02} \\

w/o local self-refinement
& 84.27
& 88.55
& 96.31
& 76.74
& 94.21
& 62.60
& 85.31
& 74.77
& 82.85
& \textbf{\tdrop{3.52}} \\

\specialrule{1.10pt}{0pt}{0pt}
\end{tabular}
\end{adjustbox}
\endgroup

\end{table*}

\begin{table*}[!t]
\centering

\begin{minipage}[t]{0.492\textwidth}
\vspace{0pt}
\centering
\captionsetup{font=footnotesize,skip=0pt}
\captionof{table}{Structural diversity analysis of topology selection on MBPP.}
\label{tab:structural_diversity}

\begingroup
\scriptsize
\setlength{\tabcolsep}{1.0pt}
\renewcommand{\arraystretch}{0.96}
\arrayrulecolor{black}

\resizebox{\linewidth}{!}{%
\begin{tabular}{
l
!{\color{black}\vrule width .45pt}c
!{\color{black}\vrule width .45pt}c
!{\color{black}\vrule width .45pt}c
!{\color{black}\vrule width .45pt}c
}
\specialrule{1.10pt}{0pt}{0pt}

\cellcolor{TabHead}\textbf{Selection Strategy}
&
\cellcolor{TabHead}\(\boldsymbol{K}\)
&
\cellcolor{TabHead}
\shortstack{
  \textbf{Active Edge}\\[-0.15ex]
  \textbf{Ratio} \(\downarrow\)
}
&
\cellcolor{TabHead}
\(\boldsymbol{D_E}\,\uparrow\)
&
\cellcolor{TabHead}
\shortstack{
  \textbf{MBPP}\\[-0.15ex]
  \textbf{Acc.} \(\uparrow\)
}
\\

\specialrule{0.75pt}{0pt}{0pt}

\rowcolor{TabStripe}
Single Best Topology
& 1
& \(12.4\;(-34.0\%)\)
& --
& \(84.74\;(-1.38)\)
\\

Quality-only Top-\(K\)
& 3
& \(18.8\;(\mathrm{ref.})\)
& \(0.41\;(\mathrm{ref.})\)
& \(86.12\;(\mathrm{ref.})\)
\\

\rowcolor{TabStripe}
Random-\(K\)
& 3
& \(18.5\;(-1.6\%)\)
& \(0.52\;(+26.8\%)\)
& \(85.37\;(-0.75)\)
\\

Max-Jaccard-\(K\)
& 3
& \(19.1\;(+1.6\%)\)
& \(\textbf{0.78}\;(+90.2\%)\)
& \(86.44\;(+0.32)\)
\\

\specialrule{0.35pt}{0pt}{0pt}

\rowcolor{OursStripe}
\textbf{\helena{} (Ours)}
& 3
& \(18.7\;(-0.5\%)\)
& \(0.69\;(+68.3\%)\)
& \(\textbf{88.73}\;(+2.61)\)
\\

\specialrule{1.10pt}{0pt}{0pt}
\end{tabular}%
}

\endgroup
\end{minipage}
\hfill
\begin{minipage}[t]{0.492\textwidth}
\vspace{0pt}
\centering
\captionsetup{font=footnotesize,skip=0pt}
\captionof{table}{Controlled noise propagation with the union graph and the
pre-injection state fixed across variants.}
\label{tab:noise_propagation}

\begingroup
\scriptsize
\setlength{\tabcolsep}{2.2pt}
\renewcommand{\arraystretch}{0.86}
\arrayrulecolor{black}

\begin{tabular}{
l
!{\color{black}\vrule width .45pt}c
!{\color{black}\vrule width .45pt}c
!{\color{black}\vrule width .45pt}c
!{\color{black}\vrule width .45pt}c
}
\specialrule{1.10pt}{0pt}{0pt}

\cellcolor{TabHead}\textbf{Variant}
&
\cellcolor{TabHead}
\shortstack{
  \textbf{Noise}\\[-0.15ex]
  \textbf{Survival} \(\downarrow\)
}
&
\cellcolor{TabHead}
\textbf{Reach@1} \(\downarrow\)
&
\cellcolor{TabHead}
\textbf{Reach@2+} \(\downarrow\)
&
\cellcolor{TabHead}
\textbf{Final Flip} \(\downarrow\)
\\

\specialrule{0.75pt}{0pt}{0pt}

\rowcolor{OursStripe}
\textbf{\helena{}}
& \textbf{34.7}
& \textbf{14.6}
& \textbf{5.3}
& \textbf{4.1}
\\

\specialrule{0.35pt}{0pt}{0pt}

\rowcolor{TabStripe}
w/o sparse activation
& 34.7
& 33.8
& 21.9
& 10.4
\\

w/o memory composer
& 100.0
& 41.6
& 17.8
& 12.7
\\

\rowcolor{TabStripe}
w/o HSC
& 100.0
& 92.4
& 68.7
& 29.6
\\

\specialrule{1.10pt}{0pt}{0pt}
\end{tabular}

\endgroup
\end{minipage}

\vspace{-2mm}
\end{table*}

\noindent\textbf{Obs.\ding{184} Union Graph Ensures Diversity.}
As shown in Table~\ref{tab:ablation}, removing the union graph causes an
average drop of \pctdown{2.88}. This result confirms its effectiveness but
does not show whether the gain comes from complementary reasoning paths or
greater communication cost. We therefore measure the structural diversity of
the selected topologies by their average pairwise edge-Jaccard distance. For
$\mathcal S=\{G_1,\ldots,G_K\}$ with edge sets $\{E_i\}_{i=1}^{K}$, it is
defined as
\begin{equation}
D_E(\mathcal S)
=
\frac{2}{K(K-1)}
\sum_{i<j}
\left(
1-
\frac{|E_i\cap E_j|}{|E_i\cup E_j|}
\right).
\end{equation}
A larger $D_E$ indicates lower structural overlap. As reported in
Table~\ref{tab:structural_diversity}, Quality-only Top-$K$ retains redundant
structures, while Max-Jaccard-$K$ achieves higher diversity but lower
accuracy. In contrast, \helena{} combines a high $D_E$ with the best MBPP
accuracy under a comparable active-edge ratio. The gain therefore comes from
reliable and nonredundant reasoning paths rather than a larger communication
budget or arbitrary diversity.

\noindent\textbf{Obs.\ding{185} Two-Level Noise Control Ensures Reliable Reasoning.}
Noise control in \helena{} requires two complementary mechanisms rather than a single step. 
Hierarchical Sparse Coordination suppresses noise during propagation by filtering private 
memories and maintaining a compact global state, preventing low-confidence traces from 
spreading through the union graph. Removing latent memory alone causes a \pctdown{3.15} drop. 
Local Self-Refinement then targets high-risk decision units and revises 
them only when contrastive evidence confirms a correction. Its removal 
triggers the largest collapse of \pctdown{3.52}, with \pctdown{5.61} on MATH-Lv5 and 
\pctdown{4.5} on MMLU-Pro. Together, the two mechanisms form a complete noise-suppression 
pipeline, with propagation control followed by targeted local revision.

\paragraph{Controlled Noise Propagation.}
We adopt controlled error injection from prior MAS robustness analysis
\citep{huang2025resilience} across the full benchmark suite. For each example,
a task-associated but incorrect intermediate record, sampled independently of
the memory selector, is inserted into a nonterminal node under a fixed union
graph and pre-injection state. $Noise Survival$ records post-filter retention;
$Reach@1$ and $Reach@2+$ measure one-hop and multi-hop exposure; $Final Flip$
counts outputs in the common clean-correct mask that become incorrect. The w/o
HSC variant disables both levels. As shown in Table~\ref{tab:noise_propagation},
the composer reduces survival to $34.7\%$; at this fixed rate, sparse activation
reduces $Reach@1/Reach@2+$ from $33.8/21.9$ to $14.6/5.3$ and yields the lowest
$Final Flip$ of $4.1\%$.

\paragraph{Sensitivity Analysis on $K$.}
Figure~\ref{fig:sensitive} shows how the number of fused topologies $K$ affects performance and cost.
$K=3$ achieves the best balance between topology diversity and redundant-edge control, so we use $K=3$ as the default setting.

\begin{figure}[t]
  \centering
  \includegraphics[width=\linewidth]{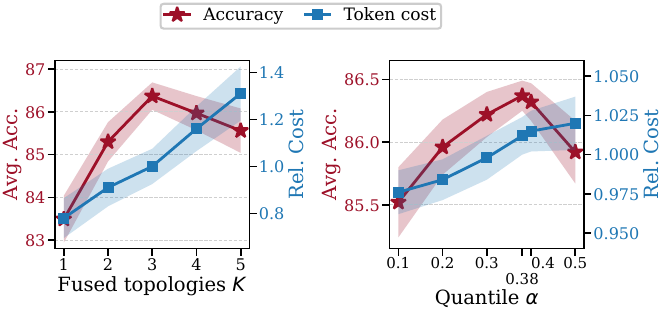}
  \caption{
  Sensitivity analysis of \helena.
  Left: performance and relative token cost as the number of fused topologies $K$ varies.
  Right: Right: performance and relative token cost as the lower-confidence quantile $\alpha$ varies.
  }
\label{fig:sensitive}
\end{figure}

\paragraph{Sensitivity Analysis on Lower-Confidence Quantile.}
Figure~\ref{fig:sensitive} also reports the effect of the lower-confidence 
quantile $\alpha$.
The default value $\alpha=0.38$ performs best, balancing uncertainty 
suppression and the retention of uncertain but useful candidate topologies.

\subsection{Case Study}

\begin{figure}[!t]
  \centering
  \includegraphics[width=\linewidth]{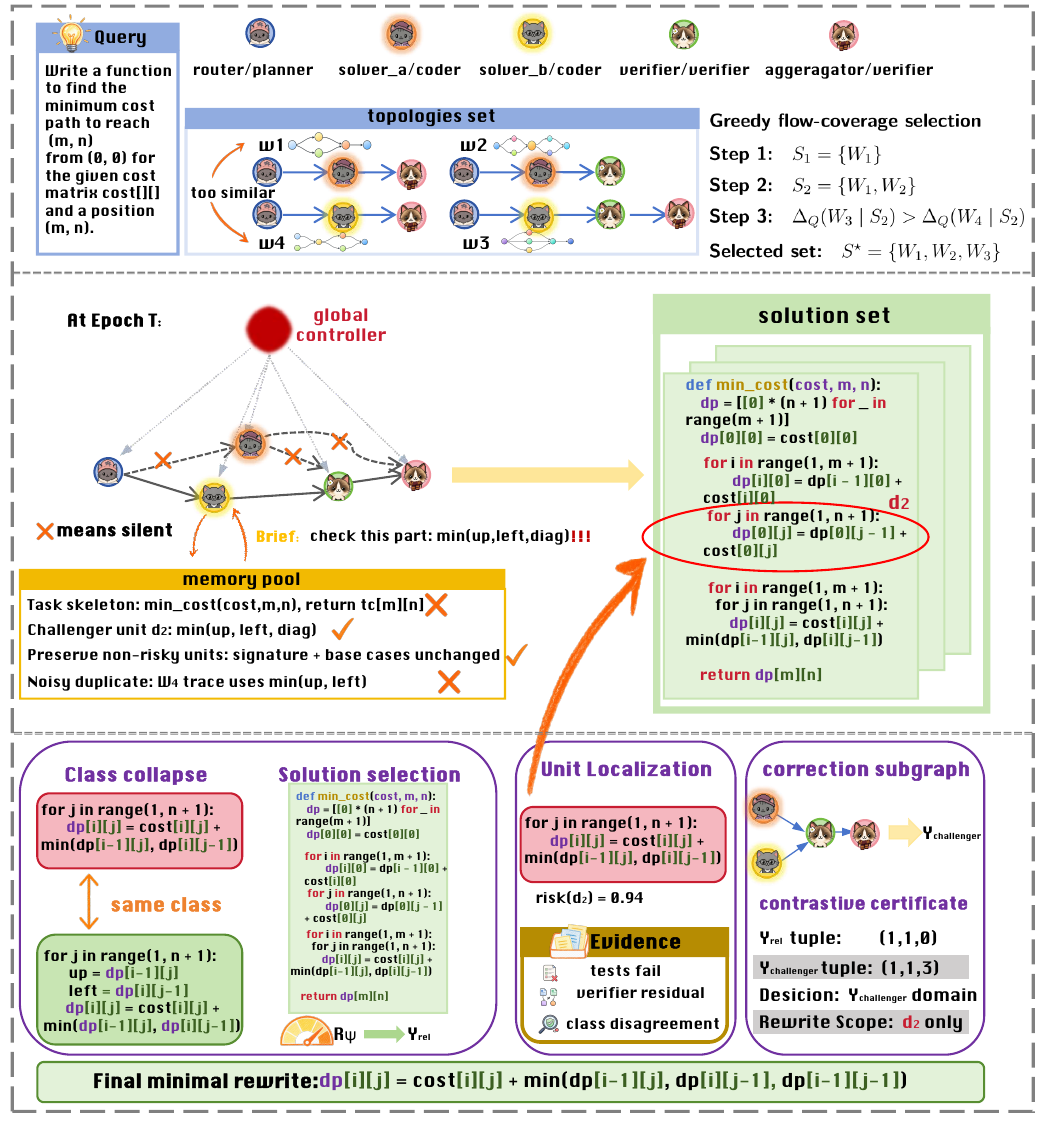}
  \caption{
  Case study illustrating the intrinsic mechanisms of \helena{}. Tuples denote \texttt{(syntax\_ok, entry\_point\_ok, passed\_tests)} in this case.
  }
  \label{fig:case}
\end{figure}

We explore and visualize the intrinsic mechanisms of \helena{}. Taking MBPP as an example, Figure~\ref{fig:case} showcases the workflow of \helena{}.

\subsection{Transferability Analysis}

Table~\ref{tab:transfer} presents cross-backbone results.
\helena{} transfers well to open-source models: Qwen3-32B remains within 1 point of GPT-4o-mini on complex reasoning benchmarks, including MATH, MATH Level 5, and MMLU-Pro, while Qwen3-8B also benefits consistently from the same coordination framework across all eight benchmarks.

\begin{table}[!t]
\centering
\caption{Cross-backbone transferability of \helena{} on Qwen3-32B and Qwen3-8B.}
\label{tab:transfer}
\begingroup
\scriptsize
\setlength{\tabcolsep}{3.5pt}
\renewcommand{\arraystretch}{1.10}
\arrayrulecolor{black}

\begin{tabular}{l|ccc}
\specialrule{1.10pt}{0pt}{0pt}

\rowcolor{TabHead}
\textbf{Benchmark}
& \textbf{GPT-4o-mini}
& \textbf{Qwen3-32B}
& \textbf{Qwen3-8B} \\

\specialrule{0.75pt}{0pt}{0pt}

\rowcolor{TabStripe}
MMLU      & \textbf{88.37} & 87.92\tdrop{0.45} & 84.12\tdrop{4.25} \\
GSM8K     & \textbf{97.39} & 96.85\tdrop{0.54} & 94.23\tdrop{3.16} \\

\rowcolor{TabStripe}
HumanEval & \textbf{95.20} & 92.75\tdrop{2.45} & 86.34\tdrop{8.86} \\
MBPP      & \textbf{88.73} & 86.90\tdrop{1.83} & 81.08\tdrop{7.65} \\

\rowcolor{TabStripe}
MATH      & \textbf{78.29} & 77.86\tdrop{0.43} & 72.84\tdrop{5.45} \\
MATH-L5   & \textbf{68.21} & 67.74\tdrop{0.47} & 61.92\tdrop{6.29} \\

\rowcolor{TabStripe}
MMLU-Pro  & \textbf{79.34} & 78.62\tdrop{0.72} & 73.41\tdrop{5.93} \\
SVAMP     & \textbf{95.48} & 94.91\tdrop{0.57} & 92.16\tdrop{3.32} \\

\specialrule{0.35pt}{1.5pt}{1.5pt}

\rowcolor{OursStripe}
\textbf{Avg.}
& \textbf{86.37}
& 85.44\tdrop{0.93}
& 80.76\tdrop{5.61} \\

\specialrule{1.10pt}{0pt}{0pt}
\end{tabular}
\endgroup
\end{table}

\FloatBarrier
\section{Conclusion}

\helena{} addresses the topology diversity and noise propagation dilemma in
LLM-based MAS 
. Across eight benchmarks, it improves the strongest baseline by
\pctup{3.47} on average and up to \pctup{10.34}.



\endgroup
\bibliography{custom}

\onecolumn
\appendix

\newenvironment{helenaalgorithm}[1]{%
  \refstepcounter{helenaalg}%
  \par\bigskip
  \noindent\rule{\linewidth}{0.8pt}\par
  \noindent\textbf{Algorithm \thehelenaalg\quad #1}\par
  \noindent\rule{\linewidth}{0.4pt}\par
  \small
  \setcounter{algline}{0}
  \begin{list}{\arabic{algline}.}{%
    \usecounter{algline}
    \leftmargin=2.2em
    \labelwidth=1.8em
    \labelsep=0.4em
    \itemsep=2pt
    \parsep=0pt
    \topsep=3pt
  }
}{%
  \end{list}
  \noindent\rule{\linewidth}{0.8pt}
  \par\bigskip
}

\newcommand{\AState}[1]{\item #1}
\newcommand{\AComment}[1]{\item[]\hspace{-1.8em}{\color{blue}\ttfamily /* #1 */}}

\section{Appendix}

\subsection{Implementation Details}
\label{app:implementation_details}

\paragraph{Model and decoding settings.}
We use \texttt{gpt-4o-mini} as the default backbone model for all agents in
\helena{} and for comparable baselines with available implementations. For
cross-backbone transfer experiments, we additionally evaluate \texttt{Qwen3-32B}
and \texttt{Qwen3-8B}. Topology descriptions, task queries, and memory records
are encoded by \texttt{Qwen3-Embedding-8B}, with embedding dimension $D=512$.
Unless otherwise specified, generation agents use temperature $0$ and
Top-$p=0.95$, while verifier and final-selection agents use deterministic
decoding with temperature $0$ and Top-$p=1.0$. Each reported result is averaged
over three independent runs.

\paragraph{Prompt and context budget.}
For each agent call, the input prompt consists of the original task query, a
role-conditioned instruction, the selected private-memory brief, active
predecessor briefs, and the current global-state summary. To control context
growth, we cap the maximum input length of each agent call at approximately
$8{,}192$ tokens and the maximum generation length at $1{,}024$ tokens. The task
query is truncated to at most $2{,}048$ tokens, the private-memory brief to
$256$ tokens, the predecessor-context brief to $384$ tokens, and the global-state
summary to $256$ tokens. When the composed prompt exceeds the budget, we first
remove zero-weight memory records and then truncate lower-weight predecessor
briefs.

\paragraph{Graph and memory settings.}
During union-graph construction, we select $K=3$ complementary topologies by
greedy $k$-DPP MAP inference and use the lower-confidence quantile
$\alpha=0.38$ for conservative topology scoring. In Hierarchical Sparse
Coordination, we run up to $T=5$ communication turns. Each active node retrieves
at most $4$ private-memory records and receives at most $3$ predecessor briefs
from active in-neighbors. Both memory retrieval and edge activation use
sparsemax, so irrelevant memory records and inactive communication edges are
assigned zero weight.

\paragraph{Stage separation.}
All learned components are optimized before benchmark evaluation and are frozen
at validation and test time. MCTS visit counts and action values are
query-local statistics and are reinitialized for every query. Test-time memory
writes are also query-local and are discarded after finalization. They do not
update model parameters or affect later test examples. The data-access protocol
is specified in Appendix~\ref{app:algorithm-workflow}.

\paragraph{Lightweight test-time topology search.}
Test-time MCTS performs legal graph edits and forward passes through the frozen
edit prior and utility predictor. It does not invoke the agent backbone or
execute candidate workflows. Agent LLM calls begin only after the selected
topologies have been merged into the query-specific union graph. The search
therefore adds no LLM token cost. Its graph operations and local model forward
passes are included in end-to-end wall-clock latency.

\begin{table}[t]
\centering
\small
\setlength{\tabcolsep}{5pt}
\begin{tabular}{l l}
\toprule
\textbf{Item} & \textbf{Setting} \\
\midrule
Default backbone & \texttt{gpt-4o-mini} \\
Transfer backbones & \texttt{Qwen3-32B}, \texttt{Qwen3-8B} \\
Embedding model & \texttt{Qwen3-Embedding-8B} \\
Embedding dimension & $512$ \\
Selected topologies & $K=3$ \\
Lower-confidence quantile & $\alpha=0.38$ \\
Maximum communication turns & $T=5$ \\
Generation temperature & $0$ \\
Verifier / finalizer temperature & $0$ \\
Max input length per call & $8{,}192$ tokens \\
Max generation length & $1{,}024$ tokens \\
Task query cap & $2{,}048$ tokens \\
Private-memory brief cap & $256$ tokens \\
Predecessor-context brief cap & $384$ tokens \\
Global-state summary cap & $256$ tokens \\
Selected memory records per node & up to $4$ \\
Active predecessor briefs per node & up to $3$ \\
Agent LLM calls during MCTS & $0$ \\
Evaluation runs & $3$ \\
Code metric & $\mathrm{pass@1}$ \\
Other metrics & exact-match accuracy \\
\bottomrule
\end{tabular}
\caption{Implementation and prompt-budget settings of \helena{}.}
\label{tab:implementation_details}
\end{table}

\subsection{Algorithm Workflow}
\label{app:algorithm-workflow}

\paragraph{Training--inference separation.}
We distinguish three disjoint data roles: a topology-supervision split
$\mathcal D_{\mathrm{sup}}$, a validation split $\mathcal D_{\mathrm{val}}$,
and the held-out test split $\mathcal D_{\mathrm{test}}$. Observed topology
utility is computed only on $\mathcal D_{\mathrm{sup}}$. Validation labels are
used only after a complete system output has been fixed, for checkpoint
selection and calibration. At test time, topology search, topology selection,
hierarchical coordination, and local refinement receive no gold answer, answer
index, exact-match result, or official held-out test feedback. The benchmark
scorer accesses such information only after the final output is irrevocably
fixed.

Prompt-visible examples or checks are treated as part of the query $Q$. They
are therefore available to every compared method and are distinct from the
official held-out evaluator. Across validation and test examples, all learned
parameters are frozen; only per-query MCTS statistics, graph activations, and
query-local memories may change.

\begin{table*}[t]
\centering
\small
\setlength{\tabcolsep}{6pt}
\begin{tabular}{lccc}
\toprule
\textbf{Signal} & \textbf{Offline supervision}
& \textbf{Validation/test inference} & \textbf{External scoring} \\
\midrule
Query and prompt-visible checks & Yes & Yes & Yes \\
Gold answer or answer index & $\mathcal D_{\mathrm{sup}}$ only & No & After output \\
Observed Tier-1/Tier-2 utility & $\mathcal D_{\mathrm{sup}}$ only & No & No \\
Frozen predicted topology utility & After fitting & Yes & No \\
Official held-out code tests & No & No & After output \\
Cross-query parameter updates & Yes & No & No \\
\bottomrule
\end{tabular}
\caption{Data-access firewall used by \helena{}. ``After output'' means that the
corresponding signal is accessed only after the system output has been fixed and
can no longer affect search, topology selection, or refinement.}
\label{tab:data_access_firewall}
\end{table*}

\paragraph{Multi-fidelity topology supervision.}
For $(Q,Y)\in\mathcal D_{\mathrm{sup}}$, executing topology $W$ produces a
candidate answer $\widehat a_W$. A training-only multi-fidelity evaluator first
uses a cheap proxy to screen all candidates, then applies Tier~1 to a shortlist
and Tier~2 to the finalists. When a higher-fidelity result is available, it
supersedes the lower-fidelity estimate rather than being linearly blended with
it. The highest-fidelity observed utility is normalized to $[0,1]$ and serves as
the supervision target:
\begin{equation}
 u^{\mathrm{obs}}(Q,W;Y)=
 \begin{cases}
 \mathbf 1[\widehat a_W=Y], & \mathrm{MCQ/EM},\\
 \mathbf 1[\mathrm{Eq}(\widehat a_W,Y)], & \mathrm{Math},\\
 \mathrm{PassRate}_{\mathcal T_{\mathrm{sup}}(Q)}(\widehat a_W),
 & \mathrm{Code}.
 \end{cases}
 \label{eq:observed_topology_utility}
\end{equation}
Here $\mathrm{Eq}$ denotes normalized numeric or symbolic equivalence, and
$\mathcal T_{\mathrm{sup}}(Q)$ contains only prompt-visible checks or checks
belonging to the non-test supervision split. Official held-out evaluator tests
are never included in $\mathcal T_{\mathrm{sup}}(Q)$. At validation and test
time, Eq.~\eqref{eq:observed_topology_utility} is not evaluated inside the
system; terminal topologies are valued only by the frozen predictor described
below.

\begin{helenaalgorithm}{Offline Topology Guidance Learning of \helena{}}

\AState{\textbf{Input:} topology-supervision set
$\mathcal D_{\mathrm{sup}}$; validation set $\mathcal D_{\mathrm{val}}$;
initial workflow $W_0$; search budget $B$; training-only multi-fidelity
evaluator $\mathcal E_{\mathrm{MF}}^{\mathrm{sup}}$; trainable edit prior
$P_\phi$; trainable utility-predictor ensemble
$\{g_{\psi_m}\}_{m=1}^{M}$.}

\AState{\textbf{Output:} frozen edit prior $P_\phi^\star$ and frozen utility
predictors $\{g_{\psi_m}^\star\}_{m=1}^{M}$.}

\AState{Initialize the topology-supervision buffer
$\mathcal D_{\mathrm{topo}}^{\mathrm{sup}}\leftarrow\emptyset$.}

\AState{\textbf{for} each $(Q,Y)\in\mathcal D_{\mathrm{sup}}$ \textbf{do}}

\AState{\quad Initialize query-local MCTS statistics
$N_Q(s,u)\leftarrow0$ and $Q_Q(s,u)\leftarrow0$.}

\AState{\quad \textbf{for} simulation $b=1,\ldots,B$ \textbf{do}}

\AState{\quad\quad Set $s\leftarrow W_0$ and search path
$\tau_b\leftarrow\emptyset$.}

\AState{\quad\quad \textbf{while} $s$ is not terminal \textbf{do}}

\AState{\quad\quad\quad Select
$u^\star\leftarrow\arg\max_{u\in\mathcal U(s)}
\mathrm{PUCT}_Q(s,u)$, where}
\[
\begin{aligned}
 \mathrm{PUCT}_Q(s,u)
 ={}&Q_Q(s,u)\\
 &+c_{\mathrm{puct}}P_\phi(u\mid s,Q)
 \frac{\sqrt{N_Q(s)}}{1+N_Q(s,u)}.
\end{aligned}
\]

\AState{\quad\quad\quad Append $(s,u^\star)$ to $\tau_b$ and set
$s\leftarrow\mathcal T(s,u^\star)$.}

\AState{\quad\quad \textbf{end while}}

\AState{\quad\quad Compile terminal state $s$ into topology $W$, execute $W$
on $Q$, and obtain}
\[
 u_W^{\mathrm{obs}}
 \leftarrow
 \mathcal E_{\mathrm{MF}}^{\mathrm{sup}}(Q,Y,W).
\]

\AState{\quad\quad Add $(Q,W,u_W^{\mathrm{obs}})$ to
$\mathcal D_{\mathrm{topo}}^{\mathrm{sup}}$.}

\AState{\quad\quad \textbf{for} each visited pair $(s,u)\in\tau_b$
\textbf{do}}

\AState{\quad\quad\quad Update the visit count}
\[
 N_Q(s,u)\leftarrow N_Q(s,u)+1.
\]
\AState{\quad\quad\quad Update the running action value}
\[
 Q_Q(s,u)\leftarrow
 Q_Q(s,u)+
 \frac{u_W^{\mathrm{obs}}-Q_Q(s,u)}{N_Q(s,u)}.
\]

\AState{\quad\quad \textbf{end for}}

\AState{\quad \textbf{end for}}

\AState{\quad Update $P_\phi$ from the utility-labelled MCTS trajectories;
only examples in $\mathcal D_{\mathrm{sup}}$ contribute to this update.}

\AState{\textbf{end for}}

\AComment{Query-conditioned topology encoding and utility learning}

\AState{\textbf{for} each $(Q,W,u_W^{\mathrm{obs}})\in\mathcal D_{\mathrm{topo}}^{\mathrm{sup}}$ \textbf{do}}

\AState{\quad Encode the query
$e_Q\leftarrow\mathrm{Emb}(Q)$.}

\AState{\quad Canonically serialize the topology as}
\[
 \mathrm{Ser}(W)=
 \{\mathrm{roles},\mathrm{edges},\mathrm{prompts}\}.
\]

\AState{\quad Compute
$e_W\leftarrow\mathrm{Emb}(\mathrm{Ser}(W))$ and
$h_W\leftarrow\mathrm{StructFeat}(W)$.}
\[
 z_W\leftarrow\mathrm{Proj}_W([e_W;h_W]).
\]
\AState{\quad Form the joint query--topology representation}
\[
 z_{Q,W}\leftarrow\mathrm{Proj}_{QW}([e_Q;z_W]).
\]

\AState{\textbf{end for}}

\AState{\textbf{for} predictor $m=1,\ldots,M$ \textbf{do}}

\AState{\quad Predict}
\[
 (\mu_{Q,W}^{(m)},\rho_{Q,W}^{(m)})
 \leftarrow g_{\psi_m}(z_{Q,W}).
\]
\AState{\quad Convert the scale parameter}
\[
 \sigma_{Q,W}^{(m)}
 \leftarrow\mathrm{softplus}(\rho_{Q,W}^{(m)})+\varepsilon.
\]

\AState{\quad Train $g_{\psi_m}$ by minimizing}
\[
 \mathcal L_{\mathrm{qual}}^{(m)}
 =
 \sum_{(Q,W,u)\in\mathcal D_{\mathrm{topo}}^{\mathrm{sup}}}
 \left[
 \frac{(u-\mu_{Q,W}^{(m)})^2}{2(\sigma_{Q,W}^{(m)})^2}
 +
 \frac12\log(\sigma_{Q,W}^{(m)})^2
 \right].
\]

\AState{\textbf{end for}}

\AState{Select and calibrate checkpoints on
$\mathcal D_{\mathrm{val}}$ only after each validation output has been fixed;
freeze all selected parameters.}

\AState{\textbf{return} $P_\phi^\star$ and
$\{g_{\psi_m}^\star\}_{m=1}^{M}$.}

\end{helenaalgorithm}

\begin{helenaalgorithm}{Test-Time Query-Conditioned Union Graph Construction of \helena{}}

\AState{\textbf{Input:} test query $Q$; initial workflow $W_0$; frozen edit
prior $P_\phi^\star$; frozen utility predictors
$\{g_{\psi_m}^\star\}_{m=1}^{M}$; search budget $B$; selected topology number
$K$; PUCT coefficient $c_{\mathrm{puct}}$; lower-confidence quantile
$\alpha<0.5$; DPP temperature $\eta$; maximum communication turns $T$.}

\AState{\textbf{Output:} union MAS graph $G_U=(V_U,E_U)$ and edge descriptors
$e_{u\to v}=(p_{uv},\mathrm{compat}_{uv},\mathrm{ctx}_{uv})$.}

\AState{No gold answer or official held-out test is an input to this algorithm.}

\AState{Initialize $\mathcal C\leftarrow\emptyset$,
$N_Q(s,u)\leftarrow0$, and $Q_Q(s,u)\leftarrow0$.}

\AComment{Candidate generation with predicted leaf values}

\AState{\textbf{for} simulation $b=1,\ldots,B$ \textbf{do}}

\AState{\quad Set $s\leftarrow W_0$ and $\tau_b\leftarrow\emptyset$.}

\AState{\quad \textbf{while} $s$ is not terminal \textbf{do}}

\AState{\quad\quad Select}
\[
u^\star
\leftarrow
\operatorname*{arg\,max}_{u\in\mathcal U(s)}
\left[
Q_Q(s,u)
+
c_{\mathrm{puct}}P_\phi^\star(u\mid s,Q)
\frac{\sqrt{N_Q(s)}}{1+N_Q(s,u)}
\right].
\]

\AState{\quad\quad Append $(s,u^\star)$ to $\tau_b$ and set
$s\leftarrow\mathcal T(s,u^\star)$.}

\AState{\quad \textbf{end while}}

\AState{\quad Compile $s$ into topology $W$ and compute $z_{Q,W}$ as in the
offline algorithm. Candidate workflows are not executed.}

\AState{\quad \textbf{for} predictor $m=1,\ldots,M$ \textbf{do}}

\AState{\quad\quad Predict
$(\mu_{Q,W}^{(m)},\sigma_{Q,W}^{(m)})
\leftarrow g_{\psi_m}^\star(z_{Q,W})$.}

\AState{\quad \textbf{end for}}

\AState{\quad Compute the predictive mean}
\[
\mu_{Q,W}
\leftarrow
\frac{1}{M}
\sum_{m=1}^{M}\mu_{Q,W}^{(m)}.
\]

\AState{\quad Compute the predictive uncertainty}
\[
\sigma_{Q,W}^{2}
\leftarrow
\frac{1}{M}
\sum_{m=1}^{M}
\left(
(\sigma_{Q,W}^{(m)})^2
+
(\mu_{Q,W}^{(m)})^2
\right)
-
\mu_{Q,W}^{2}.
\]

\AState{\quad Set
$\widehat u_\psi(Q,W)\leftarrow\mu_{Q,W}$.}

\AState{\quad \textbf{for} each $(s,u)\in\tau_b$ \textbf{do}}

\AState{\quad\quad Update}
\[
N_Q(s,u)\leftarrow N_Q(s,u)+1.
\]
\[
Q_Q(s,u)
\leftarrow
Q_Q(s,u)
+
\frac{\widehat u_\psi(Q,W)-Q_Q(s,u)}{N_Q(s,u)}.
\]

\AState{\quad \textbf{end for}}

\AState{\quad Add the unique record
$(W,\mu_{Q,W},\sigma_{Q,W})$ to $\mathcal C$.}

\AState{\textbf{end for}}

\AComment{Quality estimation and information-flow comparison}

\AState{\textbf{for} each $W\in\mathcal C$ \textbf{do}}

\AState{\quad Compute}
\[
q_{Q,W}
\leftarrow
\exp
\left(
\frac{
\mu_{Q,W}
+
\Phi^{-1}(\alpha)\sigma_{Q,W}
}{
2\eta
}
\right).
\]

\AState{\quad Map role-compatible nodes to a common order and construct the
normalized adjacency matrix $A_W$.}

\AState{\quad Compute}
\[
\Psi_T(W)
\leftarrow
\sum_{t=1}^{T}A_W^t.
\]

\AState{\textbf{end for}}

\AState{\textbf{for} each pair
$(W_a,W_b)\in\mathcal C\times\mathcal C$ \textbf{do}}

\AState{\quad Compute}
\[
\Sigma_{ab}
\leftarrow
\frac{
\left\langle
\operatorname{vec}(\Psi_T(W_a)),
\operatorname{vec}(\Psi_T(W_b))
\right\rangle
}{
\left\|
\operatorname{vec}(\Psi_T(W_a))
\right\|_2
\left\|
\operatorname{vec}(\Psi_T(W_b))
\right\|_2
}.
\]

\AState{\quad Set
$L_{ab}\leftarrow q_{Q,W_a}\Sigma_{ab}q_{Q,W_b}$.}

\AState{\textbf{end for}}

\AState{Greedily select $K$ candidates by the objective in
Eq.~\eqref{eq:dpp-selection} and denote the result by $S^\star$.}

\AComment{Union graph construction}

\AState{Align nodes by role-agent identity and construct}
\[
G_U
\leftarrow
\left(
\bigcup_{W\in S^\star}V_W,
\bigcup_{W\in S^\star}E_W
\right).
\]

\AState{\textbf{for} each edge $(u,v)\in E_U$ \textbf{do}}

\AState{\quad Compute}
\[
p_{uv}
\leftarrow
\frac{
\sum_{W\in S^\star}
q_{Q,W}\mathbf 1[(u,v)\in E_W]
}{
\sum_{W\in S^\star}q_{Q,W}
}.
\]

\AState{\quad Compute $\mathrm{compat}_{uv}$ from the role and prompt
representations of the two endpoints.}

\AState{\quad Compute $\mathrm{ctx}_{uv}$ from the finite-turn flow and the
position of the edge in the selected topologies.}

\AState{\quad Set
$e_{u\to v}\leftarrow
(p_{uv},\mathrm{compat}_{uv},\mathrm{ctx}_{uv})$.}

\AState{\textbf{end for}}

\AState{\textbf{return} $G_U$ and
$\{e_{u\to v}\mid(u,v)\in E_U\}$.}

\end{helenaalgorithm}

\paragraph{Offline training of hierarchical coordination.}
The memory query builder $f_q$, memory selector $f_s$, edge scorer
$\Psi_{\Theta_e}^{\mathrm{edge}}$, and global updater
$\mathrm{GlobalNode}_{\Theta_g}$ are optimized only on
$\mathcal D_{\mathrm{sup}}$ and frozen before validation/test inference. Memory
retrieval uses retrospective contrastive credit assignment,
\begin{equation}
 \mathcal L_{\mathrm{mem}}
 =
 -\log
 \frac{\exp z_{\mathrm{pos}}}
 {\exp z_{\mathrm{pos}}+
  \sum_{z_{\mathrm{neg}}\in\mathcal N_i}\exp z_{\mathrm{neg}}}.
 \label{eq:appendix_mem_loss}
\end{equation}
For an execution trace $\tau$, define
$s_{\Theta_e}(\tau)=
\sum_t\sum_{(v_j,v_i)\in E^t}\log\beta_{ji}^t$. The controller is trained by
pairwise trace ranking with explicit communication-cost penalties,
\begin{equation}
\begin{aligned}
 \mathcal L_{\mathrm{ctrl}}
 ={}&
 -\log
 \frac{\exp s_{\Theta_e}(\tau^+)}
 {\exp s_{\Theta_e}(\tau^+)+
  \sum_{\tau^-}\exp s_{\Theta_e}(\tau^-)}\\
 &+\lambda_E\sum_t|E^t|
 +\lambda_C\sum_t\mathrm{Tok}(D^t).
\end{aligned}
\label{eq:appendix_ctrl_loss}
\end{equation}
These are stage-specific objectives; gradients do not pass through discrete
MCTS expansion or $k$-DPP subset selection.

\begin{helenaalgorithm}{Test-Time Hierarchical Sparse Coordination of \helena{}}

\AState{\textbf{Input:} query $Q$; union graph $G_U=(V_U,E_U)$; edge
descriptors $\{e_{u\to v}\}$; private memories $\{\mathcal M_i^0\}$; role
instructions $\{r_i\}$; role-conditioned prompts $\{P_i^t\}$; turns $T$;
frozen query builder $f_q^\star$; frozen memory selector $f_s^\star$; memory
composer $\mathrm{Compose}$; frozen edge scorer
$\Psi_{\Theta_e^\star}^{\mathrm{edge}}$; frozen global updater
$\mathrm{GlobalNode}_{\Theta_g^\star}$; feedback operator
$\mathrm{Feedback}_{\mathrm{avail}}$.}

\AState{\textbf{Output:} solution set
$Y=\{(a_k,\xi_k)\}_{k=1}^{K_y}$ and activated graph trace
$G_{\mathrm{act}}^{1:T}$.}

\AState{$\mathrm{Feedback}_{\mathrm{avail}}$ uses only signals available during
inference and excludes benchmark labels.}

\AState{Initialize $g^0\leftarrow\mathrm{InitGlobal}(Q)$,
$Y\leftarrow\emptyset$, and
$G_{\mathrm{act}}^{1:T}\leftarrow\emptyset$.}

\AState{\textbf{for} turn $t=1,\ldots,T$ \textbf{do}}

\AComment{Node-level memory composition}

\AState{\quad \textbf{for} each node $v_i\in V_U$ \textbf{do}}

\AState{\quad\quad Compute
$q_i^t\leftarrow f_q^\star(Q,r_i)$.}

\AState{\quad\quad Compute
$z_{ik}^t\leftarrow f_s^\star(q_i^t,m_{ik}^t)$
for every $m_{ik}^t\in\mathcal M_i^t$.}

\AState{\quad\quad Compute}
\[
\boldsymbol\alpha_i^t
\leftarrow
\operatorname{sparsemax}
\left(
\{z_{ik}^t\mid m_{ik}^t\in\mathcal M_i^t\}
\right).
\]

\AState{\quad\quad Set}
\[
\mathcal R_i^t
\leftarrow
\{m_{ik}^t\in\mathcal M_i^t\mid\alpha_{ik}^t>0\}.
\]

\AState{\quad\quad Compute}
\[
(\ell_i^t,b_i^t,B_i^t)
\leftarrow
\operatorname{Compose}
\left(
\{(\alpha_{ik}^t,m_{ik}^t)\mid m_{ik}^t\in\mathcal R_i^t\}
\right).
\]

\AState{\quad \textbf{end for}}

\AComment{Edge-level sparse activation and execution}

\AState{\quad \textbf{for} each edge $(v_j,v_i)\in E_U$ \textbf{do}}

\AState{\quad\quad Compute}
\[
\phi_{ji}^t
\leftarrow
\Psi_{\Theta_e^\star}^{\mathrm{edge}}
\left(
b_j^t,b_i^t,g^t,e_{v_j\to v_i}
\right).
\]

\AState{\quad \textbf{end for}}

\AState{\quad \textbf{for} each target node $v_i\in V_U$ \textbf{do}}

\AState{\quad\quad Compute}
\[
\boldsymbol\beta_i^t
\leftarrow
\operatorname{sparsemax}
\left(
\{\phi_{ji}^t\mid(v_j,v_i)\in E_U\}
\right).
\]

\AState{\quad \textbf{end for}}

\AState{\quad Define}
\[
E^t
\leftarrow
\{(v_j,v_i)\in E_U\mid\beta_{ji}^t>0\}.
\]

\AState{\quad Let $V^t$ contain the endpoints of $E^t$ and set
$G^t\leftarrow(V^t,E^t)$.}

\AState{\quad Append $G^t$ to $G_{\mathrm{act}}^{1:T}$.}

\AState{\quad \textbf{for} each active node $v_i\in V^t$ \textbf{do}}

\AState{\quad\quad Form $D_i^t$ from the verbalized briefs received through
active incoming edges.}

\AState{\quad\quad Execute}
\[
o_i^t
\leftarrow
v_i
\left(
P_i^t,Q,\ell_i^t,D_i^t,g^t
\right).
\]

\AState{\quad\quad Obtain}
\[
\mathcal F_i^t
\leftarrow
\mathrm{Feedback}_{\mathrm{avail}}(o_i^t,Q).
\]

\AState{\quad\quad Update the query-local memory}
\[
\mathcal M_i^{t+1}
\leftarrow
\mathcal M_i^t
\cup
\operatorname{Write}(o_i^t,\mathcal F_i^t,G^t).
\]

\AState{\quad\quad \textbf{if} $o_i^t$ is a solution proposal
\textbf{then}}

\AState{\quad\quad\quad Create $\xi_i^t$ that links $o_i^t$ to the active
graph trace and its supporting evidence.}

\AState{\quad\quad\quad Set
$Y\leftarrow Y\cup\{(o_i^t,\xi_i^t)\}$.}

\AState{\quad\quad \textbf{end if}}

\AState{\quad \textbf{end for}}

\AState{\quad Pool the active briefs and outputs into $p^t$.}

\AState{\quad Pool the available feedback into $\mathcal F^t$.}

\AState{\quad Compute}
\[
g^{t+1}
\leftarrow
\operatorname{GlobalNode}_{\Theta_g^\star}
\left(
g^t,p^t,\mathcal F^t
\right).
\]

\AState{\textbf{end for}}

\AState{Set $K_y\leftarrow|Y|$ and discard the query-local memory writes.}

\AState{\textbf{return} $Y$ and $G_{\mathrm{act}}^{1:T}$.}

\end{helenaalgorithm}

\paragraph{Offline training of local refinement.}
The reliability estimator $R_\psi$, discrepancy encoder
$\mathrm{Disc}_\theta$, risk scorer $\mathrm{risk}_\theta$, and certificate
model $\mathrm{Cert}_\chi$ are trained only from retrospective records derived
from $\mathcal D_{\mathrm{sup}}$. They are frozen before validation and test
inference.

A reliability preference pair contains two candidates produced for the same
supervision query. The candidate with higher observed utility is preferred.
Pairs with equal utility are omitted. Let
$p=((a^+,\xi^+),(a^-,\xi^-))$ denote one pair in
$\mathcal P_{\mathrm{rel}}^{\mathrm{sup}}$. The reliability estimator is trained
by
\begin{equation}
\mathcal L_{\mathrm{rel}}
=
-\sum_{p\in\mathcal P_{\mathrm{rel}}^{\mathrm{sup}}}
\log\sigma
\left(
R_\psi(a^+,\xi^+)-R_\psi(a^-,\xi^-)
\right).
\label{eq:appendix_rel_loss}
\end{equation}

The discrepancy encoder and risk scorer form one unit risk model. For each
training unit $d$, $\mathrm{Disc}_\theta$ maps the evidence linked to the unit
to $\epsilon_d$. The risk scorer maps this representation to the risk
probability $p_d$. The loss below updates both parts of the model. No separate
discrepancy loss is used.
\begin{equation}
\begin{aligned}
p_d
&=
\sigma(\mathrm{risk}_\theta(d;\epsilon_d)),\\
\mathcal L_{\mathrm{risk}}
&=
-\sum_{(d,\epsilon_d,r_d^\star)\in
\mathcal D_{\mathrm{risk}}^{\mathrm{sup}}}
\left[
r_d^\star\log p_d
+
(1-r_d^\star)\log(1-p_d)
\right].
\end{aligned}
\label{eq:appendix_risk_loss}
\end{equation}
The target $r_d^\star$ indicates whether the unit lacks sufficient support
under the supervision available in $\mathcal D_{\mathrm{sup}}$.

For each proposed replacement, the certificate model receives the challenger,
the current solution, and the evidence linked to both versions of the unit.
Its output and training loss are
\begin{equation}
\begin{aligned}
c_d
&=
\mathrm{Cert}_\chi
\left(
y_d^{\mathrm{ch}},
\xi_d^{\mathrm{ch}},
y_{\mathrm{rel}},
\xi_{\mathrm{rel}},
d
\right),\\
\mathcal L_{\mathrm{cert}}
&=
-\sum_{(d,y_d^{\mathrm{ch}},\xi_d^{\mathrm{ch}},
 y_{\mathrm{rel}},\xi_{\mathrm{rel}},c_d^\star)
\in\mathcal D_{\mathrm{cert}}^{\mathrm{sup}}}
\left[
c_d^\star\log c_d
+
(1-c_d^\star)\log(1-c_d)
\right].
\end{aligned}
\label{eq:appendix_cert_loss}
\end{equation}
A positive target is assigned only when the local replacement improves the
supervised outcome of unit $d$ while preserving the remaining units. No
validation or test label is used to update these models.

\paragraph{Local refinement operators.}
The operator $\mathrm{Collapse}$ follows semantic answer equivalence
\citep{kuhn2023semantic}. Given the resulting answer classes,
$\mathrm{Rep}$ retains the candidate with the highest $R_\psi$ score from each
class.

\begin{table}[t]
\centering
\small
\setlength{\tabcolsep}{5pt}
\begin{tabularx}{\linewidth}{l X}
\toprule
\textbf{Task type} & \textbf{Decision unit returned by $\mathrm{UnitParse}$} \\
\midrule
Multiple choice & One normalized answer choice \\
Mathematical reasoning & One derivation step or the final answer field \\
Code generation & One code span identified by the available verifier evidence \\
\bottomrule
\end{tabularx}
\caption{Task-specific decision units used by Local Self-Refinement.}
\label{tab:decision_units}
\end{table}

For a selected solution $(y_{\mathrm{rel}},\xi_{\mathrm{rel}})$,
$\mathrm{UnitParse}$ returns the locally verifiable units listed in
Table~\ref{tab:decision_units}. For each unit $d$, $\mathrm{Disc}_\theta$
receives the unit and the evidence linked to it. It produces the
fixed-dimensional representation $\epsilon_d$, which is passed to
$\mathrm{risk}_\theta$ and is not itself a risk score.

The local score $R_\psi^{\mathrm{loc}}(a,\xi,d)$ uses the same reliability
estimator after evidence unrelated to unit $d$ has been masked. The global
score and local score therefore share parameters.

The certificate model compares the current unit with a proposed replacement
using their linked evidence. Its positive output supports only the local
replacement and does not certify the full solution.

The operator $\mathrm{ExtractSubgraph}$ starts from the nodes cited by the
provenance of unit $d$. It follows the recorded active paths to the nodes that
supplied evidence for the unit. Verifier nodes that assessed the unit are
retained. The union graph defines eligible edges, while the activated trace
limits the correction graph to paths used in the current execution.

The operator $\mathrm{RepairExec}$ executes only the agents in the correction
graph. Its prompt contains the original query and the selected solution. The
disputed unit and its linked evidence form the repair context. The prompt asks
for a replacement of that unit while preserving the remaining solution. Each
challenger is returned with a new evidence record.

\noindent Let
$\Theta_{\mathrm{ref}}^\star=
(R_\psi^\star,\mathrm{Disc}_\theta^\star,
 \mathrm{risk}_\theta^\star,\mathrm{Cert}_\chi^\star)$
denote the frozen local refinement modules.

\begin{helenaalgorithm}{Test-Time Local Self-Refinement of \helena{}}

\AState{\textbf{Input:} query $Q$; solution set
$Y=\{(a_k,\xi_k)\}_{k=1}^{K_y}$; union graph $G_U$; activated trace
$G_{\mathrm{act}}^{1:T}$; frozen module bundle
$\Theta_{\mathrm{ref}}^\star$; risk threshold $\tau$; margin $\delta$.}

\AState{\textbf{Output:} final reliable solution $\widehat y$.}

\AState{All evidence used below is available during inference. No gold answer
or official held-out test is accessible.}

\AComment{Decision Unit Localization}

\AState{Collapse semantically equivalent answers
$\mathcal G\leftarrow\mathrm{Collapse}(Y)$.}

\AState{\textbf{for} each equivalence class $G\in\mathcal G$ \textbf{do}}

\AState{\quad Select}
\[
(a_G,\xi_G)
\leftarrow
\operatorname*{arg\,max}_{(a,\xi)\in G}
R_\psi^\star(a,\xi).
\]

\AState{\textbf{end for}}

\AState{Set $\mathcal R\leftarrow\mathrm{Rep}(\mathcal G)$ and select}
\[
(y_{\mathrm{rel}},\xi_{\mathrm{rel}})
\leftarrow
\operatorname*{arg\,max}_{(a,\xi)\in\mathcal R}
R_\psi^\star(a,\xi).
\]

\AState{Parse the selected solution}
\[
\mathcal D_{\mathrm{unit}}
\leftarrow
\mathrm{UnitParse}(y_{\mathrm{rel}},\xi_{\mathrm{rel}}).
\]

\AState{Initialize $\mathcal U\leftarrow\emptyset$.}

\AState{\textbf{for} each $d\in\mathcal D_{\mathrm{unit}}$ \textbf{do}}

\AState{\quad Compute}
\[
\epsilon_d
\leftarrow
\mathrm{Disc}_\theta^\star
(d,y_{\mathrm{rel}},\xi_{\mathrm{rel}},Y).
\]

\AState{\quad Compute
$r_d\leftarrow\mathrm{risk}_\theta^\star(d;\epsilon_d)$.}

\AState{\quad \textbf{if} $r_d>\tau$ \textbf{then}
$\mathcal U\leftarrow\mathcal U\cup\{d\}$.
\textbf{end if}}

\AState{\textbf{end for}}

\AComment{Adversarial Solution Validation}

\AState{\textbf{for} each $d\in\mathcal U$ \textbf{do}}

\AState{\quad Extract}
\[
G_d^{\mathrm{corr}}
\leftarrow
\mathrm{ExtractSubgraph}
(G_U,G_{\mathrm{act}}^{1:T},d,\xi_{\mathrm{rel}},Y).
\]

\AState{\quad Retrieve}
\[
\mathcal H_d^{\mathrm{ret}}
\leftarrow
\{(a,\xi)\in Y\mid\mathrm{Disagree}(a,y_{\mathrm{rel}},d)=1\}.
\]

\AState{\quad Generate}
\[
\mathcal H_d^{\mathrm{gen}}
\leftarrow
\mathrm{RepairExec}
(Q,G_d^{\mathrm{corr}},y_{\mathrm{rel}},d,\xi_{\mathrm{rel}}).
\]

\AState{\quad Set
$\mathcal H_d\leftarrow
\mathcal H_d^{\mathrm{ret}}\cup\mathcal H_d^{\mathrm{gen}}$
and $\mathcal A_d\leftarrow\emptyset$.}

\AState{\quad \textbf{for} each
$(y_d^{\mathrm{ch}},\xi_d^{\mathrm{ch}})\in\mathcal H_d$ \textbf{do}}

\AState{\quad\quad Compute}
\[
\Delta_d
\leftarrow
R_\psi^{\star,\mathrm{loc}}
(y_d^{\mathrm{ch}},\xi_d^{\mathrm{ch}},d)
-
R_\psi^{\star,\mathrm{loc}}
(y_{\mathrm{rel}},\xi_{\mathrm{rel}},d).
\]

\AState{\quad\quad Compute}
\[
c_d
\leftarrow
\mathrm{Cert}_\chi^\star
\left(
y_d^{\mathrm{ch}},
\xi_d^{\mathrm{ch}},
y_{\mathrm{rel}},
\xi_{\mathrm{rel}},
d
\right).
\]

\AState{\quad\quad \textbf{if} $\Delta_d>\delta$ and $c_d=1$
\textbf{then}}

\AState{\quad\quad\quad Set
$\mathcal A_d\leftarrow
\mathcal A_d\cup\{(y_d^{\mathrm{ch}},\xi_d^{\mathrm{ch}})\}$.}

\AState{\quad\quad \textbf{end if}}

\AState{\quad \textbf{end for}}

\AState{\quad \textbf{if} $\mathcal A_d\neq\emptyset$ \textbf{then}}

\AState{\quad\quad Select}
\[
(y_d^\star,\xi_d^\star)
\leftarrow
\operatorname*{arg\,max}_{(y,\xi)\in\mathcal A_d}
R_\psi^{\star,\mathrm{loc}}(y,\xi,d).
\]

\AState{\quad\quad Rewrite}
\[
y_{\mathrm{rel}}
\leftarrow
\mathrm{Rewrite}(y_{\mathrm{rel}},d,y_d^\star).
\]

\AState{\quad\quad Update}
\[
\xi_{\mathrm{rel}}
\leftarrow
\mathrm{UpdateEvidence}
(\xi_{\mathrm{rel}},d,\xi_d^\star,G_d^{\mathrm{corr}}).
\]

\AState{\quad \textbf{else} preserve unit $d$ unchanged.
\textbf{end if}}

\AState{\textbf{end for}}

\AState{Set $\widehat y\leftarrow y_{\mathrm{rel}}$.}

\AState{\textbf{return} $\widehat y$.}

\end{helenaalgorithm}

\paragraph{External benchmark scoring.}
After Test-Time Local Self-Refinement returns $\widehat y$, the system state is
frozen and no further search, topology selection, or rewriting is permitted.
Only then does the external benchmark scorer compare $\widehat y$ with the gold
answer or run the official held-out tests. Consequently, benchmark feedback
cannot affect any output-producing decision.

\subsection{Controlled Noise Propagation Protocol}
\label{app:noise_propagation}

The evaluation covers the full benchmark suite under a shared injection
protocol. In each run, a task-associated but incorrect intermediate record is
sampled independently of the memory-selection scores and inserted into a
nonterminal node. Thus, the injected record is related to the task but is not
constructed to rank among the node's selected memory records. For every
injection run, the query, injected record, injection source, union graph, and
pre-injection state are identical across variants. Local Self-Refinement is
disabled throughout this experiment.

Let
\(
\mathcal A=
\{\text{HELENA},\text{w/o sparse activation},
\text{w/o memory composer},\text{w/o HSC}\}
\)
denote the evaluated variants. \helena{} enables both HSC levels;
w/o sparse activation retains the memory composer but executes all union edges;
w/o memory composer retains sparse edge activation but bypasses memory
selection; and w/o HSC disables both levels, yielding dense union-graph
execution with the full private memory available to each active node.

Let $\mathcal I$ denote the common set of injection runs, and let
$\widetilde m_i$ be the record injected in run $i$. For variant $a\in\mathcal A$,
let $\mathcal R_i^a$ denote the records exposed after node-level memory
selection; when the memory composer is disabled, $\mathcal R_i^a$ is the full
private memory and therefore contains $\widetilde m_i$. Let
$\mathcal V_i^{(1)}$ contain nodes whose shortest-path distance from the
injection source is one, and let $\mathcal V_i^{(2+)}$ contain nodes at distance
$2$ through the maximum communication horizon $T$. Because the union graph and
injection source are fixed, these eligible node sets are identical across
variants.

The indicator $x_{iv}^a$ equals one when the provenance of
$\widetilde m_i$ appears in the input context of node $v$ under variant $a$.
A node that is not activated, or is activated without receiving the injected
record, has $x_{iv}^a=0$. Reach is micro-averaged over all eligible downstream
nodes, so every variant uses the same denominators. The reported metrics are

\begin{align}
\mathrm{NoiseSurvival}^{a}
&=
\frac{100}{|\mathcal I|}
\sum_{i\in\mathcal I}
\mathbf 1[\widetilde m_i\in\mathcal R_i^{a}],
\label{eq:noise_survival}\\
\mathrm{Reach@1}^{a}
&=
100
\frac{
\sum_{i\in\mathcal I}
\sum_{v\in\mathcal V_i^{(1)}}x_{iv}^{a}
}{
\sum_{i\in\mathcal I}
|\mathcal V_i^{(1)}|
},
\label{eq:reach_one}\\
\mathrm{Reach@2+}^{a}
&=
100
\frac{
\sum_{i\in\mathcal I}
\sum_{v\in\mathcal V_i^{(2+)}}x_{iv}^{a}
}{
\sum_{i\in\mathcal I}
|\mathcal V_i^{(2+)}|
}.
\label{eq:reach_two_plus}
\end{align}

Let $\widehat y_{i,a}^{\mathrm{clean}}$ and
$\widehat y_{i,a}^{\mathrm{inj}}$ denote the outputs of variant $a$ before and
after injection, and let $y_i$ denote the reference answer. To ensure a common
denominator, Final Flip is evaluated on the shared clean-correct mask
\begin{equation}
\mathcal C
=
\left\{
 i\in\mathcal I
 \mid
 \widehat y_{i,a}^{\mathrm{clean}}=y_i
 \text{ for every }a\in\mathcal A
\right\}.
\label{eq:common_clean_mask}
\end{equation}
Thus, the same examples contribute to every row of Table~\ref{tab:noise_propagation}:
\begin{equation}
\mathrm{FinalFlip}^{a}
=
\frac{100}{|\mathcal C|}
\sum_{i\in\mathcal C}
\mathbf 1[
\widehat y_{i,a}^{\mathrm{inj}}\neq y_i
].
\label{eq:final_flip}
\end{equation}

\subsection{Prompt Set}
\label{app:prompt-set}

This appendix lists representative prompt schemas used by \helena{}.
We include prompts for topology-level prompt slots, role-conditioned agents,
latent-guided graph execution, deductive reasoning contracts, and local
verification / repair. At validation and test time, all placeholders are
populated only from the input query, frozen model outputs, graph traces,
query-local memories, and prompt-visible checks. They never contain the gold
answer, answer index, or official held-out evaluator tests.

\begin{promptblock}{Discrete Prompt Slot Template}{PromptGreen}
reasoning_mode = [
  "Answer directly with minimal reasoning.",
  "Reason step by step before giving the answer.",
  "First critique possible failure modes, then answer."
]

upstream_usage = [
  "Use upstream outputs only when they provide direct evidence.",
  "Summarize useful upstream information before reasoning.",
  "Ignore upstream information if it conflicts with the task constraints."
]

output_style = [
  "Return a concise natural-language answer.",
  "Return bullet points followed by a final answer.",
  "Return a compact structured response with stable fields."
]

verification_mode = [
  "No explicit verification.",
  "Perform a short final consistency check.",
  "Check the answer against every task constraint."
]

finalization = [
  "Answer only.",
  "Answer followed by a short rationale.",
  "Final answer must appear after the marker FINAL:"
]
\end{promptblock}

\begin{promptblock}{Representative Agent Role Prompts}{PromptPurple}
planner_system_prompt = """
You are a planning agent. Understand the problem first and provide a concise,
reliable plan for solving it. Focus on decomposing the task into useful steps.
"""

reasoner_system_prompt = """
You are a reasoning agent. Produce a complete and traceable reasoning process.
Avoid skipping important intermediate steps.
"""

coder_system_prompt = """
You are a coding agent. Translate the problem into executable algorithmic steps
or code-level reasoning. Preserve required function names and interfaces.
"""

verifier_system_prompt = """
You are a verifier agent. Check whether a candidate answer satisfies the task
constraints, output format, and available evidence. Report only grounded issues.
Do not assume access to a gold answer or hidden evaluator tests.
"""

summarizer_system_prompt = """
You are a summarizer agent. Synthesize multiple candidates into a concise,
consistent final proposal. Do not introduce unsupported claims.
"""
\end{promptblock}

\begin{promptblock}{Latent-Guided Runtime Prompt}{PromptBlue}
runtime_user_prompt = """
Question:
{rendered_question}

Latent-guided memory brief:
{memory_brief}

Current task:
Turn {t}/{T}. {role_instruction}

Use your private memory together with graph-mediated neighbour signals.
Only rely on neighbour information if it is relevant to your current role.
Avoid repeating irrelevant traces.
Do not assume access to a gold/reference answer or hidden evaluator tests.

Task context:
{task_context}

Output contract:
{answer_contract}
"""
\end{promptblock}

\begin{promptblock}{Deductive Solver and Verifier Contracts}{PromptOrange}
deductive_solver_contract = """
SOLUTION:
1. <quantity_name> = <arithmetic expression> = <value>
2. <quantity_name> = <arithmetic expression> = <value>
3. ...

FINAL: <number_or_expression>
"""

deductive_verifier_contract = """
VERDICT: pass|challenge|uncertain
LOCUS: step=<id>|final|parse
ISSUE: arithmetic_mismatch|algebra_mismatch|unsupported_transition|missing_final|none
FIX: <one sentence>
"""
\end{promptblock}

\begin{promptblock}{Local Verification and Repair Prompts}{PromptRed}
anchor_verification_probe = """
You are verifying a proposed final answer.

Problem:
{question}

Proposed final answer:
{reliable_solution}

Do not assume the proposed answer is correct.
Re-solve the problem from the task quantities only.
Use a compact equation-chain format.
Every step must contain a computable equation.
Do not use a gold/reference answer or hidden evaluator tests.

Return exactly:

SOLUTION:
1. <quantity_name> = <arithmetic expression> = <value>
2. <quantity_name> = <arithmetic expression> = <value>
...

FINAL: <answer>
"""

suffix_repair_prompt = """
You are repairing a derivation.

Problem:
{question}

Verified prefix. Do not change these steps:
{verified_prefix}

First inconsistent step:
{bad_step}

Verifier residual:
{residual}

Rewrite only from Step {verified_prefix_len + 1} onward.
Keep the verified prefix unchanged.
Do not use a gold/reference answer or hidden evaluator tests.

Return:

SOLUTION:
1. ...
2. ...
...

FINAL: <answer>
"""
\end{promptblock}

\end{document}